\documentclass[12pt]{iopart}
\usepackage{iopams}
\usepackage{graphicx}
\usepackage[square,numbers,sort&compress]{natbib}
\usepackage[usenames]{color}
\definecolor{violet}{rgb}{0.6,0.0,0.3}
\usepackage[%
pdftitle={Cosmology with the Large Synoptic Survey Telescope},%
pdfauthor={Zhan \& Tyson},%
bookmarksnumbered=true,%
linkcolor=blue,citecolor=blue,urlcolor=blue,colorlinks=true]{hyperref}
\usepackage{hypernat}

\def\aap{Astron.~\& Astrophys.}
\def\aj{Astron.~J.}
\def\apj{Astrophys.~J.}
\def\apjl{Astrophys.~J.~Lett.}
\def\apjs{Astrophys.~J.~Supp.}
\def\araa{Ann.~Rev.~Astron.~\& Astrophys.}
\def\nat{Nature}
\def\mnras{Mon.~Not.~Roy.~Astron.~Soc.}
\def\physrep{Phys.~Rep.}
\def\prd{Phys.~Rev.~D}

\def\zp{Zeitschrift f\"{u}r Phys.}
\def\jcap{J.~Cosmo.~Astropart.~Phys.}
\def\pasp{Pub.~Astron.~Soc.~Pacific}
\def\pasj{Pub.~Astron.~Soc.~Japan}
\def\procspie{Proc.~SPIE}
\def\aspc{ASP Conf.~Series}
\def\ssr{Space Sci.~Rev.}

\def\da{d_\mathrm{A}}
\def\Da{D_\mathrm{A}}
\def\dl{d_\mathrm{L}}
\def\Dc{D_\mathrm{C}}
\def\Om{\Omega_\mathrm{m}}
\def\Ob{\Omega_\mathrm{b}}
\def\Ok{\Omega_\mathrm{k}}
\def\Odm{\Omega_\mathrm{dm}}

\def\Ode{\Omega_\mathrm{de}}
\def\om{\omega_\mathrm{m}}
\def\ob{\omega_\mathrm{b}}
\def\odm{\omega_\mathrm{dm}}
\def\rhom{\rho_\mathrm{m}}
\def\rdm{\rho_\mathrm{dm}}
\def\GN{\mathcal{G}}
\def\rmd{\mathrm{d}}
\def\lcdm{$\Lambda$CDM}
\def\wcdm{$w$CDM}
\def\phz{{photo-{\it z}}}
\def\Phz{{Photo-{\it z}}}

\def\mpc{\,\mbox{Mpc}}
\def\mpci{\,\mbox{Mpc}^{-1}}
\def\himpc{\,h^{-1}\mbox{Mpc}}
\def\hmpci{\,h\mbox{Mpc}^{-1}}
\def\ev{\,\mbox{eV}}
\def\nm{\,\mbox{nm}}
\def\um{\,\mu\mbox{m}}
\def\m{\,\mbox{m}}
\def\sec{\,\mbox{s}}
\def\K{\,\mbox{K}}
\def\deg{\,\mbox{deg}}
\def\arcmin{\,\mbox{arcmin}}
\def\arcsec{\,\mbox{arcsec}}
\def\fnl{f_\mathrm{NL}}
\newcommand{\Msun}{\,h^{-1}\mbox{M}_\odot}

\def\newblock{\hskip .11em plus .33em minus .07em}

\newcommand{\beq}{\begin{equation}}
\newcommand{\eeq}{\end{equation}}
\newcommand{\bea}{\begin{eqnarray}}
\newcommand{\eea}{\end{eqnarray}}

\newcommand{\bit}[1]{\mathbf{#1}}
\newcommand{\remove}[1]{}

\begin{document}

\title[Cosmology with LSST]{Cosmology with the Large Synoptic 
Survey Telescope: an Overview}

\author{Hu Zhan$^1$ and J.~Anthony Tyson$^2$}

\address{$^1$CAS Key Laboratory of Space Astronomy and Technology, 
National Astronomical Observatories, A20 Datun Road, Chaoyang District, 
Beijing 100012, China\\
$^2$Department of Physics, University of California, One Shields 
Avenue, Davis, CA 95616, USA}

\ead{\mailto{zhanhu@nao.cas.cn}, \mailto{tyson@physics.ucdavis.edu}}

\begin{abstract}
The Large Synoptic Survey Telescope (LSST) is a high \'etendue imaging 
facility that is being constructed atop Cerro Pach\'{o}n in Northern 
Chile. It is scheduled to begin science operations in 2022. 
With an $8.4\m$ ($6.5\m$ effective) aperture, a novel 
three-mirror design achieving a seeing-limited $9.6\deg^2$ field of view, 
and a 3.2 Gigapixel camera, the LSST has the 
deep-wide-fast imaging capability necessary to carry out
an $18,000\deg^2$ survey in six passbands ($ugrizy$) to a coadded 
depth of $r\sim 27.5$ over 10 years using $90\%$ of its observational 
time. The remaining $10\%$ of time will be devoted to considerably deeper 
and faster time-domain observations and smaller surveys. In total, each patch of the sky
in the main survey will receive 800 visits allocated across the 
six passbands with $30\sec$ exposure visits.

The huge volume of high-quality LSST data will provide a wide 
range of science opportunities and, in particular, open a new era of 
precision cosmology with unprecedented statistical power and tight 
control of systematic errors. In this review, we give a brief account 
of the LSST cosmology program with an emphasis on dark energy 
investigations. The LSST will address dark energy physics and cosmology 
in general by exploiting diverse precision probes including  
large-scale structure, weak lensing, type Ia supernovae, galaxy clusters, 
and strong lensing. Combined with the cosmic microwave background data, 
these probes form interlocking tests on the cosmological model 
and the nature of dark energy in the presence of various systematics.

The LSST data products will be made available to the U.S. and Chilean 
scientific communities and to international partners with no proprietary 
period. Close collaborations with contemporaneous imaging and 
spectroscopy surveys observing at a variety of wavelengths, 
resolutions, depths, and timescales will be a 
vital part of the LSST science program, which will not only enhance
specific studies but, more importantly, also allow a more complete 
understanding of the universe through different windows. 

\end{abstract}

\maketitle

\tableofcontents

\section{Introduction}

Breakthrough discoveries have greatly expanded the boundary of our
perceptible universe from the solar system to the Milky Way, to the 
\emph{Realm of the Nebulae} \cite{hubble1936}, and to the afterglow of the
Big Bang \cite{penzias1965}. Tantalizing evidence for new physics from our 
cosmic quest calls for a new generation of powerful survey facilities. 
Indeed, not only do astronomers fully endorse research on the physics of 
the universe with long-term planning exercises such as \emph{Cosmic Vision} 
\cite{cosmic_vision} --- the European Space Agency's science program for 
2015-2025 and \emph{New Worlds, New Horizons in Astronomy and Astrophysics} 
\cite{astro2010} --- the 2010 US Decadal Survey of Astronomy and 
Astrophysics (Astro2010), but physicists also fully recognize the 
potential to advance our fundamental understanding of the particle world 
through its connection with the cosmos \cite{quarks_cosmos,quantum_universe}. 
What is truly exciting is that, after more than a decade of community 
efforts, ambitious projects like the Large Synoptic Survey 
Telescope\footnote{http://www.lsst.org/} (LSST) are now on track for 
operations starting in the early 2020s. 

It was realized through early dark matter mapping experiments in the 
1980s and 1990s
\cite[e.g.,][]{tyson1984,schneider1988,tyson1990,kneib1993,fahlman1994} 
that a huge survey volume was ultimately needed for useful cosmological 
tests with imaging surveys. 
Specifically, the survey should be both deep and wide. This  
enables precision cosmology by boosting the sample size for methods 
based on properties of individual objects, by suppressing the sample 
variance error and shot noise for methods relying on spatial statistics of 
the objects, and by providing information about the evolution of the 
universe. To complete such a survey in a reasonable time, one must resort
to a facility with high \'etendue or ``throughput'', which may be 
quantified by the product of the light-collecting area and the field of 
view (FoV) in solid angle. It would be ideally a 
large-aperture wide-field telescope that delivers superb image quality,
which brings considerable challenges to both the optics and the camera.

The need for a high-throughput instrument for dark matter mapping 
motivated the Big Throughput Camera 
(BTC, community access started in 1996) \cite{tyson1992, wittman1998}. 
It was hosted by the National Optical Astronomy Observatory (NOAO) at 
the 4-meter telescope (later named Blanco telescope) at 
Cerro Tololo Inter-American Observatory. 
The BTC on the Blanco telescope had the most powerful sky survey 
capability at that time. The pixels critically sampled the sub-arcsecond 
seeing, and shift-and-stare imaging provided very deep images over 
moderately wide FoV. 
Two groups of astronomers used the BTC to search for type Ia supernovae 
(SNe Ia), trying to measure the expected ``deceleration'' of the expansion 
of the universe. What they found was remarkable \cite{riess98,perlmutter99}: 
instead of decelerating, the universe is accelerating! 

While the BTC could survey to $r\sim 26$ over a few square degrees in a week, 
another facility using a larger mosaic of the same 2K$\times$2K pixel CCDs, 
but on a much smaller telescope, was optimized for moderate depth, very wide 
field observations: the Sloan Digital Sky Survey (SDSS, science operations 
started in 2000) \cite{york2000}. The first phase of the SDSS completed in
2005, and its imaging covered $8000\deg^2$ of the sky in 5 bands to 
a depth of $r \le 22.2$ \cite{sdss_dr5}. 
The SDSS has been hugely successful, broadly impacting not only 
astronomy but also the way research in astronomy is done. With three
extensions, it has now gone well beyond its original goals.

A natural question was then whether we could meet the challenges of 
surveying much deeper and, at the same time, significantly wider than 
the SDSS. An affirmative answer in the late 1990s would seem optimistic, 
but progress has been made 
by precursors such as the Deep Lens Survey \cite{dls}, 
the Canada-France-Hawaii Telescope Legacy Survey \cite{cfhtls}, 
the Panoramic Survey Telescope and Rapid Response System (Pan-STARRS)
\cite{panstarrs}, and the Dark Energy Survey \cite{des}.
With a sufficiently large \'etendue one would not have to make choices 
between wide-shallow or deep-narrow surveys; one could have a facility 
that would do the best of both: \emph{deep and wide}. Moreover, with a 
large telescope the exposures could be short, enabling
a deep-wide-fast survey which could provide the data needed by a broad 
range of science programs from a single comprehensive set of observations.
Thus was born the idea of the ``Dark Matter Telescope''
\cite{dmt_tyson2001}. 

Plans for the wide-field Dark Matter Telescope and its camera were 
presented at a workshop on gravity at SLAC National Accelerator 
Laboratory in August 1998 \cite{tyson1998}. 
The science case for such a telescope was 
submitted to the 2000 US Decadal Survey of Astronomy and Astrophysics 
(Astro2000) \cite{astro2000} in June 1999. This proposal emphasized the broad
science reach from cosmology to the time domain through data mining of a
single deep-wide-fast sky survey.  Astro2000 recommended it
highly as a facility to discover near-Earth asteroids as well as to study 
dark matter and renamed it Large-aperture Synoptic Survey Telescope 
(the word ``aperture'' is now omitted). 
Soon after, a summer workshop on wide field astronomy was organized at 
the Aspen Center for Physics in July 2001. This was the beginning of 
wide involvement by the scientific community in the LSST.  In 2002 NOAO 
set up a national committee to develop the LSST 
design reference mission \cite{strauss2004}. 
Meanwhile, plans for a multi-gigapixel focal 
plane and initial designs for the telescope-camera-data system were
being developed \cite{tyson2001,starr2002,tyson2002}.

In 2002 the National Science Foundation (NSF) funded research and 
development of the new CCDs required for the LSST, 
supplementing an investment already made by Bell Labs. L.~Seppala
modified R.~Angel's original three-mirror optical design for the telescope 
\cite{angel2000}, creating a $10\deg^2$, very low distortion FoV.  The LSST
Corporation was formed in 2002 to manage the project. An R\&D  
proposal was submitted to the NSF in early 2007 and favorably reviewed 
later that year. Thanks to a gift from C.~Simonyi and B.~Gates,  
the LSST 8.4-m primary-tertiary mirror was cast in 2008, and in early 2009 
the secondary mirror blank was cast as well. 

A grass-roots effort in 2008 and 2009 by the astronomy community 
resulted in the \emph{LSST Science Book} \cite{lsst_sb}, a 596 page compendium 
of breakthrough science applications co-authored by 245 scientists.
In addition, the community wrote many white papers on LSST science applications as
input to the Astro2010 decadal survey process.
The effort was well received, and the LSST was ranked by Astro2010 
as the highest priority for ground-based astronomy \cite{astro2010}.
In February 2011 a construction proposal was submitted to the NSF.
After many project reviews, the NSF National Science Board gave approval 
to begin LSST construction in August 2014. The LSST construction is on 
schedule, and first light for engineering tests with a commissioning camera 
is scheduled for 2019, with the decade-long main survey beginning in 2022.

With its high \'etendue and image quality, the LSST is 
naturally a powerful facility for dark energy studies, and it 
will take advantage of multiple probes such as weak lensing (WL), 
large-scale structure (LSS), 
SNe Ia, galaxy clusters, and strong lensing. It is therefore classified 
as a Stage IV dark energy experiment, i.e., a next generation project, 
by the Dark Energy Task Force (DETF) \cite{detf}. These same survey data
also enable related investigations such as the dark matter distribution 
on a variety of scales from galaxies to the LSS and the sum of
neutrino masses.

The rest of the paper is arranged as follows. \Sref{sec:framework} and
\sref{sec:frontiers} introduce, respectively, a set of frequently 
encountered concepts in cosmology and several research areas that are 
expected to advance significantly with the LSST. We give a concise 
description of the LSST project in \sref{sec:lsst} and discuss in 
\sref{sec:probes} the cosmological probes that have been developed
extensively within LSST Science Collaborations. 
Although an emphasis is given to applications for dark energy studies, 
readers are reminded that these probes are sensitive to other elements 
of cosmology as well. For more thorough discussions of the LSST --- its
design, capabilities, and wide range of science opportunities, see 
\cite{lsst_sb}; a shorter overview is also available on the 
arXiv \cite{ivezic2008}. A detailed plan of the dark energy program for 
the LSST can be found in the white paper of the LSST Dark Energy 
Science Collaboration (DESC) \cite{desc_wp}.

\section{Cosmological Framework} \label{sec:framework}
Modern cosmology has established a successful framework that enables
precision interpretation of observational data. Such a framework is built 
upon Einstein's General Relativity (GR) and the profound principle that 
the universe must be homogeneous and isotropic on sufficiently large scales.
Clearly, alternative theories of cosmology may be constructed by 
modifying either GR or the cosmological principle, which are indeed 
areas of active studies. Nevertheless, GR plus the cosmological 
principle remains the most effective and self-consistent theory to 
date that describes the universe from its very early stage to the present.
This section is hence focused on key elements of the conventional 
cosmological framework.

\subsection{Cosmic distances} \label{sec:flrw}
The Friedmann-Lema{\^i}tre-Robertson-Walker (FLRW) metric 
\cite{friedmann22,lemaitre27,robertson35,walker35}
is essential for cosmology under the principle of spatial homogeneity 
and isotropy. The line element of the FLRW metric is given by
\beq \label{eq:flrw}
\rmd s^2 = -c^2 \rmd t^2 + a^2(t)
\left(\frac{\rmd r^2}{1-Kr^2}+r^2\rmd\Omega\right),
\eeq
where $a(t)$ and $K$ are the scale factor and the curvature of the 
universe, respectively. For convenience, the scale factor $a(t)$ is 
set to unity at present time $t_0$. The FLRW metric is independent of GR. 
The influence of gravity is through the dynamics, e.g., the evolution of 
$a(t)$, so quantities and relations derived solely from the FLRW metric 
are often applicable to models of alternative gravity theories. 

It is useful to define a distance that depends on the spatial coordinates 
only:
\beq \label{eq:dcS}
\rmd \Dc^2 = \frac{\rmd r^2}{1-Kr^2}+r^2\rmd\Omega.
\eeq
Since the distance $\Dc$ between two bodies at rest locally in the 
expanding universe does not change with time, it is given the name 
``comoving distance.'' For a photon heading toward the observer
($\rmd s = \rmd \Omega = 0$), the radial comoving distance along 
its geodesic is
\beq \label{eq:dc}
\Dc[r(t)] = \int_{0}^{r} \frac{\rmd r'}{\sqrt{1-Kr'^2}} = 
\int_{t}^{t_0} \frac{c \rmd t'}{a(t')}.
\eeq
With \eref{eq:dc}, one can link 
the observed redshift $z$ of spectral lines of a distant object to the 
scale factor $a$ at the time of emission via $1+z = a^{-1}$ 
\cite{lemaitre27}. The comoving distance between the 
observer and the object at $z$ can then be written as
\beq \label{eq:dcz}
\Dc(z)=\int_{0}^{z}\frac{c\rmd z'}{H(z')},
\eeq
where the Hubble parameter $H[z(t)] \equiv \dot{a}/a$
is a measure of the expansion rate of the universe at $z$ (or $t$),
and the over-dot on $a$ denotes derivative with respect to $t$. 
For two objects along the line of sight at $z_1$ and $z_2$ ($z_2 > z_1$), 
respectively, the comoving distance between them  $\Dc(z_1,z_2)$ equals 
$\Dc(z_2)-\Dc(z_1)$. For completeness, the time since Big Bang is given 
by
\beq
t(z) = \int_{z}^{\infty} \frac{\rmd z'}{(1+z') H(z')}.
\eeq

Two practical definitions of distances are frequently used on cosmic 
scales: the angular diameter distance $\da$ and the luminosity distance
$\dl$. The former is the distance that converts the angular size 
$\theta$ of an object into its linear size $l$ perpendicular to the 
line of sight, i.e., $l \simeq \da \theta$  ($|\theta| \ll 1$). 
The luminosity distance is 
the radius of the sphere at which an isotropic light source with 
luminosity $L$ would produce the observed flux $F$, i.e., $L=4\pi\dl^2F$. 
By setting $dt = dr = 0$ in \eref{eq:flrw}, one can see that $a r$ 
is just the angular diameter distance of the coordinate $r$ as viewed 
from the origin. The integral over $r$ in \eref{eq:dc} can be carried 
out to get
\beq \label{eq:SDc}
r \equiv S_\mathrm{K}(\Dc) = \left\{ 
\begin{array}{ll} 
K^{-1/2} \sin \left[K^{1/2} \Dc(z) \right] & K > 0 \\
\Dc(z) & K = 0 \\
(-K)^{-1/2} \sinh \left[(-K)^{1/2} \Dc(z) \right] & K < 0
\end{array}
\right . ,
\eeq
so that 
\beq \label{eq:da}
\Da(z) \equiv (1+z) \da(z) = S_\mathrm{K}\left[\Dc(z)\right],
\eeq
where $\Da(z)$ is the comoving angular diameter distance.
In lensing studies, one often needs to calculate the angular diameter 
distance of an object at $z_2$ as viewed by an observer at $z_1$ 
($z_2 > z_1$):
\beq
\Da(z_1,z_2) \equiv (1+z_2) \da(z_1,z_2) = 
S_\mathrm{K}\left[\Dc(z_1,z_2)\right].
\eeq
The luminosity distance and the angular diameter distance
satisfy the reciprocity relation \cite{etherington33} (also known as 
the distance-duality relation), i.e., $\dl(z) = (1+z)^2 \da(z)$.
It is valid as long as \cite[e.g.,][]{bassett04,uzan04}: 
(1) space-time is described by a metric theory, 
(2) photon geodesics are unique, and (3) photons are
neither created nor destroyed along the geodesics. 
The reciprocity relation is therefore independent of the FLRW metric or 
GR, and it offers a fairly model 
independent test of cosmology.

\subsection{Growth of perturbations} \label{sec:exp-grow}

Gravity drives the evolution of the universe. Hence, we put GR into 
context here. From Einstein's field equations one can derive the 
Friedmann equations
\bea \label{eq:adot}
\left(\frac{\dot{a}}{a}\right)^2 = 
\frac{8\pi \GN}{3}(\rho_\mathrm{m}+\rho_\mathrm{de}) - 
\frac{Kc^2}{a^2} \\ \label{eq:addot}
\frac{\ddot{a}}{a} = -\frac{4\pi \GN}{3}
\left[\rho_\mathrm{m} + \rho_\mathrm{de} (1 + 3 w_\mathrm{de})\right],
\eea
where $\GN$ is the gravitational constant, $\rho_\mathrm{m}$ is the matter
density\footnote{Including both ordinary matter and dark matter. 
The former is often referred to as baryons in astronomy for convenience.}, 
$\rho_\mathrm{de}$ is the dark energy density, $w_\mathrm{de}$ is the dark 
energy equation of state (EoS), and we have neglected the radiation
component as well as the pressure of matter. 
\Eref{eq:addot} shows that $w_\mathrm{de} < -1/3$ is 
a necessary (but not sufficient) condition for dark energy to drive the 
cosmic acceleration ($\ddot{a}>0$). Conservation of energy leads to a 
generic scaling 
\beq
\rho_\mathrm{x}(z) = \rho_\mathrm{x}(0) \exp\left[3 \int_0^z 
\frac{1+w_\mathrm{x}(z')}{1+z'} \rmd z'\right], 
\eeq
where the subscript x can be ``m'' or ``de.'' Since the EoS of matter 
$w_\mathrm{m}$ is practically zero in the redshift range directly observed 
by the LSST, we can drop the subscript ``de'' in $w_\mathrm{de}$ without
confusion. \Eref{eq:adot} can now be rewritten as 
\beq \label{eq:h2}
\frac{H^2(z)}{H_0^2}=\Om (1+z)^3 + \Ok (1+z)^2 + 
\Ode \exp\left[3 \int_0^z \frac{1+w(z')}{1+z'} \rmd z'\right],
\eeq
where $H_0 \equiv H(0)$ is the Hubble constant, 
$\Om \equiv \rho_\mathrm{m}(0)/ \rho_\mathrm{c}$ with 
$\rho_\mathrm{c} \equiv 3H_0^2/(8\pi \mathcal{G})$, 
$\Ok \equiv -K c^2/H_0^2$, and 
$\Ode \equiv \rho_\mathrm{de}(0)/\rho_\mathrm{c} = 1 - \Om - \Ok$. 
If dark energy is just the cosmological constant $\Lambda$ 
($w=-1$), $\rho_\mathrm{de}$ indeed will be a constant over time.
A flat universe dominated by the cosmological constant and 
cold dark matter (CDM) is often referred to as the \lcdm{} universe, 
and we use \wcdm{} to denote the case $w \neq -1$.
\Eref{eq:h2} shows that $H_0$, $\Om$, $\Ode$ (or $\Ok$), and $w(z)$ 
completely specify the cosmic expansion history, which, in turn, 
determines the distances and time defined in \sref{sec:flrw}. 

Gravitational instability turns minute initial fluctuations into 
structures we observe today. The growth history of these 
fluctuations provides crucial cross-checks of the 
cosmological model. On linear scales, the overdensity of the perturbed
density field $\rho_\mathrm{m}(\bi{x},t)$ can be decomposed into a spatial 
component and a time component
\beq
\delta_\mathrm{m}(\bi{x},t) \equiv 
\frac{\rho_\mathrm{m}(\bi{x},t)-\rho_\mathrm{m}(t)}{\rho_\mathrm{m}(t)}
 = \delta_\mathrm{m}(\bi{x}) G(t),
\eeq
where  $G(t)$ is the linear growth function. Assuming that dark energy 
does not cluster on scales of interest, one obtains for
the fluctuations in matter
\beq \label{eq:G}
\ddot{G} + 2H \dot{G} = 4\pi\mathcal{G}\rho_\mathrm{m} G.
\eeq
It can be 
seen that the Hubble expansion in \eref{eq:G} works against gravity, 
so the cosmic acceleration slows down the growth of structures. 
Gravity in general affects both the Hubble expansion and the 
right-hand side of \eref{eq:G}. Therefore, one can potentially 
distinguish dark energy from modified gravity 
theories by examining both the expansion history (or distances) and 
the growth history of the universe 
\cite[e.g.,][]{linder05,knox2006b,ishak2006}
\emph{if} dark energy is completely homogeneous and isotropic. 

In analyses of galaxy redshift surveys, one often needs the 
logarithmic growth rate
\beq \label{eq:f}
f(z) = \frac{\rmd \ln G}{\rmd \ln a} \approx \left[\Om(z)\right]^\gamma,
\eeq
where the growth index $\gamma \sim 0.55$--$0.6$ is not overly sensitive 
to cosmological parameters within the GR framework 
\cite{peebles80,lightman90,lahav91,wang98}. 

\subsection{Two-point statistics of fluctuations} \label{sec:stats}

Statistics of the cosmic density field are crucial probes of the 
universe. In the linear regime, the cosmic density 
field can be approximated by a Gaussian random field, whose 
properties are all captured in its two-point statistics, i.e., the 
correlation function in configuration space or, equivalently, the 
power spectrum in Fourier space. For this reason, we discuss only the 
two-point statistics here. For higher-order statistics and their 
applications, see, e.g., \cite{white79,szapudi93,takada03}.

The correlation function of the overdensity $\delta(\bi{x})$
is defined as
\beq \label{eq:xi}
\xi(\Delta x) = \langle \delta(\bi{x}) \delta(\bi{x}')\rangle,
\eeq
where $\langle\ldots\rangle$ denotes an ensemble average, and, because
of isotropy, the correlation function depends only on the 
separation $\Delta x = |\bi{x}' - \bi{x}|$  between two points. 
Under the assumption of homogeneity and ergodicity, one can conveniently 
replace the ensemble average in \eref{eq:xi} with a volume average. 
Real observations are made on the past light-cone, not on a snapshot 
(i.e., a constant-time hypersurface) of the universe, so the volume 
average of the light-cone differs slightly from that of the snapshot,
which is seen in \emph{N}-body simulations 
\cite{wagner08,kim09}. Such an effect is deterministic and can be 
precisely calibrated.

The distribution of galaxies may differ from that of matter. A clustering bias is
thus introduced to account for the difference between the galaxy 
correlation function $\xi_\mathrm{g}$ and the matter correlation
function, i.e., $\xi_\mathrm{g} = b^2 \xi$. The galaxy bias evolves with 
time and depends on the halo mass \cite{blanton00,
somerville01,weinberg04,tinker10}. 
It also varies with the scale but changes rather slowly above tens 
of $\himpc$ \cite[e.g.,][]{manera10,wang13}. While the galaxy 
bias is a complex subject of research \cite[see][for 
early theoretical investigations]{cole89,mo96,catelan98,sheth99b},
we treat it as a constant for simplicity. In analyses of galaxy 
clustering data, it is useful to model the galaxy bias in detail, 
so that one can extract cosmological information from small scales 
\cite{hu04a,zheng07,cacciato12}.

The correlation of the Fourier modes 
$\hat{\delta}(\bi{k})$ defines the power spectrum $P(k)$
\beq \label{eq:ps}
 \langle \hat{\delta}(\bi{k})\hat{\delta}^*(\bi{k}')\rangle
= (2\pi)^3\delta^\mathrm{D}(\bi{k}-\bi{k}') P(k),
\eeq
where $\delta^\mathrm{D}(\bi{k}-\bi{k}')$ is the Dirac delta function.
The power spectrum is often expressed in a dimensionless form
$\Delta^2(k) = k^3P(k)/2\pi^2$, which is roughly the amplitude of 
fluctuations in the logarithmic interval around $k$.
By Fourier expanding $\delta(\bi{x})$ in \eref{eq:xi}, one finds 
that the correlation function in configuration space is just the 
Fourier transform of $P(k)$, i.e.,
\beq 
\xi(\Delta x) = \frac{1}{(2\pi)^3} \int 
\e^{i\bi{k}\cdot\bi{x}} P(k) \rmd^3 k.
\eeq
Real surveys have finite volume and resolution, so one 
applies the discrete Fourier transform in practice. 
Since the power spectrum and the correlation function are equivalent,
data analyses can be performed with either statistic. Still, the 
complexity of the analyses depends on both the adopted statistic 
and the application. 
Besides the multiplicative galaxy bias, the galaxy power spectrum 
receives an additive term due to the shot noise 
\beq \label{eq:gps}
P_\mathrm{g}(k) = b^2 P(k) + n_\mathrm{g}^{-1},
\end{equation}
where $n_\mathrm{g}$ is the galaxy number density. The galaxy bias in 
\eref{eq:gps} is equivalent to that in the galaxy correlation function 
\emph{if} it is scale-independent.

The statistical error of the power spectrum at the wavevector 
$\bi{k}$ equals the power spectrum itself, i.e., 
$\sigma_{P_\mathrm{g}}(\bi{k}) = P_\mathrm{g}(k)$ \cite{feldman94}.
The errors at different wavevectors are independent under a
Gaussian approximation, so the uncertainty of a band power 
can be reduced effectively
\beq
\sigma_{P_\mathrm{g}}(k) = \sqrt{\frac{2}{N_k}} P_\mathrm{g}(k),
\eeq
where $N_k$ is the number of modes within a band of width $\Delta k$. 
For a survey of volume $V$, $N_k \simeq k^2\Delta kV/2\pi^2$.
The uncorrelated errors make the power spectrum convenient for 
at least theoretical studies. 
With observational effects and nonlinearity, one can 
still decorrelate the modes with some effort 
\cite{vogeley1996,hamilton00}.

In the linear regime, where $\Delta^2(k) \ll 1$, 
all the modes grow at the same rate with 
no coupling to each other. Therefore, the linear power spectrum 
at a redshift well below the redshift of the cosmic microwave 
background (CMB, $z\sim 1100$) can be scaled from that at a 
reference redshift (e.g., $z=0$) using the linear growth 
factor
\beq
P_\mathrm{L}(k,z) = \frac{G^2(z)}{G^2(0)} P_\mathrm{L}(k,0).
\eeq
Because of the complex nature of the galaxy bias, it is not 
straightforward to predict the galaxy power spectrum precisely. 
One usually has to fit the galaxy bias parameter(s) with data. 
However, it is encouraging that relatively simple bias models based 
on the concept of halos \cite{sheth99b,peacock00,scoccimarro01} are 
largely consistent with current observations 
\cite{marinoni05,seljak05,blanton06,marin13}. 

Perturbative calculations can extend the power spectrum prediction 
into the weakly nonlinear regime \cite{crocce06,jeong06,mcdonald07,
matarrese07,matsubara08,pietroni08,taruya12}. Going further, one 
must resort to cosmological \emph{N}-body simulations. 
Tests with simulations have reached $1\%$ level accuracy out to 
$k \sim 1 \hmpci$ \cite{heitmann10b}, which is within a factor of a few 
from the requirements of future surveys \cite{huterer05b,hearin12}. 
However, the cost of running \emph{N}-body simulations makes them 
impractical for direct use in cosmological parameter estimation. 
A solution to this problem
is in essence to develop an advanced interpolation scheme that can 
quickly output the power spectrum with satisfactory accuracy from
a minimum set of simulations spanning the parameter space 
\cite{heitmann09,heitmann10a,agarwal12,agarwal14}. For less 
demanding applications, fitting formulae for the nonlinear matter
power spectrum \cite{peacock96,smith03,takahashi12} are convenient 
to use.

\section{Cosmic Frontiers} \label{sec:frontiers}

Cosmology is a key science driver of the LSST. Although great discoveries 
often come unexpected, there are many areas for which one can predict
substantial progress with the LSST. Here we give a brief account of 
several such areas.

\subsection{Accelerating universe}

The cosmic acceleration is undoubtedly a profound challenge to our 
understanding of the universe \cite{quarks_cosmos,quantum_universe,detf}. 
So far investigations have been focused on two classes of models: 
dark energy and modified gravity 
\cite[for recent reviews, see][]{caldwell2009,clifton12,koyama2016}. 
The former is developed under the framework in \sref{sec:framework} as 
a special component of the universe, while the latter induces the 
acceleration with a new form of gravity. There are also models that 
admit no new components or physics but attribute the acceleration to a 
breakdown of the cosmological principle or an oversimplification of GR 
effects in the real universe 
\cite[e.g.,][]{buchert1997,clarkson2011,buchert2012}. 
These models are less popular, but they do invite a closer inspection 
of the cosmological framework and the evidence for the cosmic acceleration.

Consider SNe Ia as an example. 
One may fit their luminosity distances with a
Friedmann model in \sref{sec:exp-grow}, and the acceleration can be 
deduced from the resulting model parameters. 
Because the result is obtained under a model that allows acceleration
in the first place, it is hard to draw a definitive conclusion before 
exhausting all other possibilities. Alternatively, 
one can estimate the deceleration parameter --- a kinematic quantity
\beq \label{eq:q}
q = -\frac{a^2}{\dot{a}^2}\frac{\ddot{a}}{a} = 
\frac{\rmd \ln H}{\rmd \ln (1+z)} - 1
\eeq
without referring to the dynamics \cite{robertson55,hoyle56}. Studies 
along this line indeed show that $q < 0$, i.e., $\ddot{a} > 0$, at low 
redshift \cite{turner02,shapiro06,neben13}. Since $\ddot{a}$ involves
the second derivative with respect to time, measurements of the cosmic 
time or age of the universe at a series of redshifts would give the 
most direct evidence. 
However, such measurements are rather challenging. For example,
the redshift drift effect can map the evolution of the cosmic expansion 
rate in theory \cite{sandage62,mcvittie62}. But even if one 
could monitor an object's redshift stably over decades to measure 
a $\sim 10^{-9}$ change in its redshift, peculiar velocities and other 
uncertainties associated with the object itself could easily dominate 
over the signal \cite{linder97,loeb98}. A more practical example is 
to measure the cosmic expansion rate from age differences of passively 
evolving galaxies \cite{jimenez02}. This technique has achieved typical
precision of $5$-$14\%$ on the Hubble parameter up to $z \sim 1$
\cite{moresco12}, though further improvement is needed to quantify 
the acceleration precisely via \eref{eq:q}.

A negative deceleration parameter signifies acceleration in an FLRW 
universe. If, however, the cosmological principle does not hold, then 
one cannot use a global scale factor like that in the FLRW metric to 
describe the expansion of the universe. Consequently, the deceleration 
parameter, if measurable at all, becomes a local quantity, and a 
local acceleration does not necessarily mean a global acceleration of 
the universe. This is the case for models which place the observers 
in an accelerating void \cite{tomita01,alnes06,biswas07}, which are
however in tension with the data \cite{moss11,zhang11}. 
In addition, nonlinear evolution of small-scale inhomogeneities have
been postulated to cause an apparent acceleration on much larger 
scales \cite{buchert1997,buchert00,rasanen04,ellis05,kolb2006}.
Although such an effect is found to be negligible 
\cite{siegel05,kasai06,baumann12}, the question about the validity of the 
FLRW metric for interpreting real observations is highly relevant and 
deserves careful examination.

Applying the FLRW metric with GR, one finds that a
smooth component with strong negative pressure, i.e., dark energy,
is needed to drive the accelerated expansion. 
Its simplest form, the cosmological constant $\Lambda$, was invented by 
Einstein nearly a century ago to keep the universe static. 
Decades before the SN Ia results in 1998, many had already argued 
for a positive cosmological constant based on a range of observations 
including the LSS, CMB, Hubble constant, and so on 
\cite{gunn75,peebles84,efstathiou90,ostriker95,krauss95}. Even 
a time-varying cosmological ``constant'' due to a scalar field
was proposed by Peebles \& Ratra in 1988 \cite{peebles88}. 
Depending on the behavior of the scalar field's 
kinetic and potential terms, one arrives at
the quintessence model ($-1<w<1$) \cite{zlatev99}, the phantom model
($w<-1$) \cite{chiba00}, and the quintom model ($w$ can cross $-1$ 
with the help of two fields) \cite{feng05}. Many more dark energy 
models have been discussed in the literature, and interested readers
are referred to \cite{silvestri09} for a review.

Phenomenologically, dark energy is characterized by its present energy 
density as a fraction of the critical density $\Ode$ and its EoS $w$.
A major task of dark energy experiments is thus to measure $\Ode$ and 
reconstruct $w$ as a function of redshift for model comparison.
A widely used parametrization of the EoS is $w = w_0 + w_a(1-a)$ 
\cite{chevallier01,linder03a}. The reciprocal of the area of the 
$w_0$-$w_a$ error ellipse was introduced as a figure of merit by the 
DETF \cite{detf} to evaluate the performance of various 
surveys\footnote{The report uses the 95\% confidence limit, i.e., 
roughly 2$\sigma$ in the Gaussian case, to define the error ellipse, but 
studies afterward frequently use the 68.3\% confidence limit (1$\sigma$) 
instead.}. To fully utilize the capability of future surveys
(so-called Stage IV dark energy experiments), one should go beyond the 
$w_0$-$w_a$ parametrization \cite{albrecht07,albrecht2009}.

Besides strong negative pressure, dark energy might also have a sound 
speed that is sufficiently low to allow appreciable clustering on very 
large scales. 
The sound speed of standard quintessence is equal to the speed of light, 
while models such as \emph{k}-essence \cite{armendariz-picon00} 
can produce a sound speed well below the speed of light 
over certain period of time \cite{erickson02}. The effect of dark energy 
clustering might be detected with CMB and galaxy surveys on very 
large scales \cite{bean04,hu04b}. 

Modified gravity offers another mechanism to drive the cosmic 
acceleration. There are two well studied models: 
Dvali-Gabadadze-Porrati (DGP) gravity \cite{dvali00} 
and $f(R)$ gravity \cite{capozziello2002,carroll04}. 
In the DGP model, matter is confined in a 4-dimensional brane, while 
gravity can leak into the fifth dimension above a transition scale 
$r_\mathrm{c}$, causing it to weaken faster than expected in a 
4-dimensional space-time. The $f(R)$ model replaces the Ricci scalar 
$R$ in the GR gravitational action with a function $f(R)$. A suitable 
choice of $f(R)$ could accelerate the cosmic expansion.
Unfortunately, neither model appears viable. 
On the one hand, DGP gravity is inconsistent with 
observations \cite{song07}. On the other hand, $f(R)$ gravity is
constrained to be so close to GR that dark energy is still needed
to drive the acceleration \cite{jain2013}. 
Nonetheless, it is useful to see from a specific example how to 
generate the accelerated expansion. Hence, we include a few equations 
of the DGP model here. The Friedmann equation becomes
\cite{deffayet01}
\beq \label{eq:dgp:friedmann}
H^2 - \epsilon \frac{c}{r_\mathrm{c}}\sqrt{H^2+\frac{K c^2}{a^2}} = 
\frac{8\pi\mathcal{G}}{3}\rho_\mathrm{m} - \frac{K c^2}{a^2},
\eeq
where $\epsilon = \pm 1$. An acceleration in the DGP 
model is produced with $\epsilon = +1$.
The linear growth function satisfies \cite{lue06}
\beq \label{eq:dgp:grow}
\ddot{G} + 2H\dot{G} = 4\pi\GN\left(1+\frac{1}{3\beta}\right)\rhom G,
\eeq
where
\beq\label{eq:dgp:beta}
\beta=1-2\frac{Hr_\mathrm{c}}{c}\left(1+\frac{\dot{H}}{3H^2}\right).
\eeq
Comparing with the corresponding equations in GR, one sees that 
DGP gravity (and modified gravity in general) affects the linear growth
by altering both the expansion background, i.e., $H$ on the left-hand
side of \eref{eq:dgp:grow}, and the effective strength of gravity on 
the right-hand side.

Distinguishing dark energy from modified gravity is of particular 
interest, as the physics behind them are fundamentally different.
If dark energy only affects the background expansion, then 
one may detect the signature of modified gravity from inconsistency
between the expansion history and the growth history. With more generic 
(parametrizations of) dark energy and modified gravity models, the task
is thought to be impossible \cite{kunz07,bertschinger08}, though it 
may still be feasible for dark energy models with no coupling to 
matter \cite{jain08}.

\begin{figure}
\centering
\includegraphics[width=0.85\columnwidth]{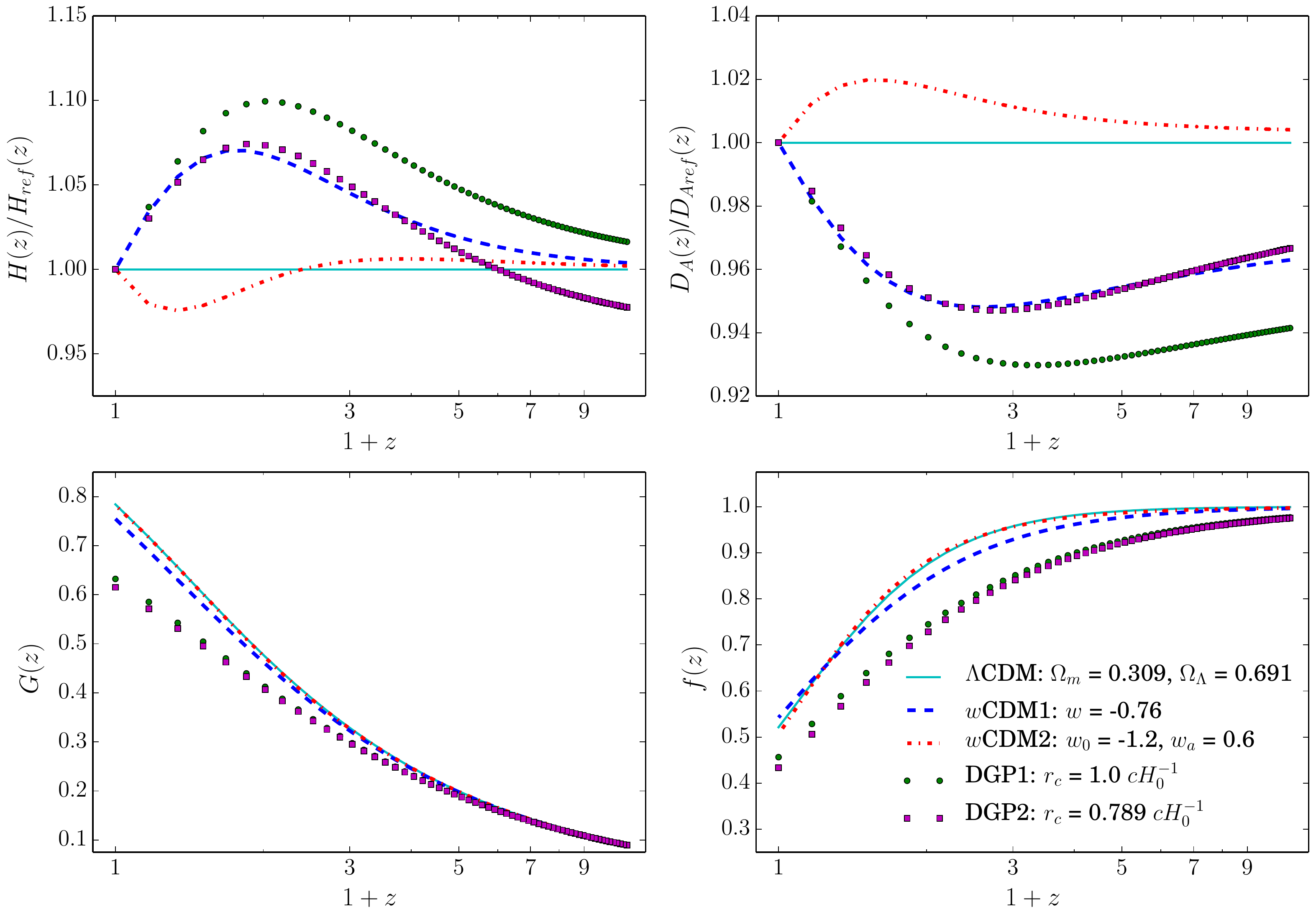}
\caption{Upper left panel: Hubble parameter of various models relative to
that of the \lcdm{} model with $\Om=0.309$ (solid line). 
The models are \wcdm{}1 with $w=-0.8$ (dashed line), \wcdm{}2 with 
$w_0=-1.2$ and $w_a=0.6$ (dot-dashed line), DGP1 with $r_\mathrm{c}=c H_0^{-1}$ 
(circles), and DGP2 with $r_\mathrm{c}=0.789c H_0^{-1}$ (squares). All models
are flat. Upper right panel: same as the upper left panel but for the 
angular diameter distance. Lower left panel: linear growth factor of 
all the models. Lower right panel: growth rate of all the models.
\label{fig:hdgf}}
\end{figure}

Redshift evolution of the Hubble parameter, angular diameter distance,
linear growth function, and growth rate are shown in 
\fref{fig:hdgf} for five cosmological models. 
The parameters of the \lcdm{} model adopt \emph{Planck} 2015 results
\cite{planck2015par}. The \wcdm1 and \wcdm2 models differ from the \lcdm{} 
model only in the dark energy EoS. It is relatively easy to tune 
different models to match either the expansion or the growth history, 
but not so easy to match both. For instance, whereas the \wcdm{}1 model 
is practically indistinguishable from the DGP2 model in terms of $\Da(z)$ 
and $H(z)$, the differences in $G(z)$ and $f(z)$ are conspicuous. Although 
the \wcdm{}2 model coincides with the \lcdm{} model in $G(z)$ and $f(z)$,
measurements of $\Da(z)$ and $H(z)$ to better than $1\%$ at multiple 
redshifts below $z \sim 2$ can tell them apart. A similar case occurs 
between the DGP1 and the DGP2 models. 

\Fref{fig:hdgf} clearly demonstrates the value of accurately mapping both 
the expansion history and the growth history of the universe. 
The Hubble parameter and different types of distances allow us to 
distinguish the \lcdm{} model from dynamical dark energy models, 
while the growth function and growth rate are useful for breaking the 
degeneracy between modifications to gravity and the background expansion 
effect. Note that spectroscopic surveys are needed to measure $H(z)$ and
$f(z)$ from radial baryon acoustic oscillations (BAOs) 
\cite{hu03b,linder03b,seo03,gaztanaga2009,sanchez2013}
and the redshift distortion effect 
\cite{kaiser1987,peacock2001,guzzo2008,alam2017}, respectively. 
Future imaging and spectroscopic surveys  will 
be able to constrain the quantities in \fref{fig:hdgf} to the percent 
level in many redshift bins up to 
$z \lesssim 3$ \cite{zhan2009a,schlegel11,amendola13}, which is sufficient 
to distinguish models with even smaller differences than those shown. 

\subsection{Dark matter} \label{sec:dm}

Dark matter is another major frontier of cosmology and particle physics and 
has a much longer history of study than dark energy. The existence of dark 
matter is evidence for physics beyond the standard model. Astrophysical 
evidence for the existence of dark matter comes from many directions 
including, for example, dynamics of galaxy clusters, galaxy rotation curves, 
X-ray emissions from galaxies and clusters, and WL mass mapping 
\cite{zwicky1933,rubin1978,faber1979,fabricant1980,kriss1983,clowe2004}. 
Formation of galaxies and substructures within them requires dark matter, 
or at least the main component of it, to have a low velocity dispersion 
(hence ``cold'') in the early universe \cite{blumenthal1984}. 
Out of the diverse topics in dark matter research, we can only touch 
upon a few that are relevant to the LSST. 
Readers are referred to \cite{bertone2005,feng2010} for 
thorough reviews.

Optical observations are crucial to dark matter studies that probe
its gravitational effects. An important application is to determine 
the mean density parameter of dark matter $\odm = \Odm h^2$ (equivalent 
to $\rdm$; ditto matter density $\om$ and baryon density $\ob$)
or its fraction $\Odm$ in the total matter-energy budget of the 
universe. One can estimate $\Odm$ in a number of ways. 
Distance measurements can constrain the total matter fraction $\Om$, 
which is the sum of $\Odm$ and $\Ob$ (neglecting radiation and other 
minor components), and then $\Odm$ can 
be obtained with a prior on $\Ob$, which might also be deduced from 
the same survey data. Sensitivity of the abundance of massive halos and 
its evolution to $\Om$ (in combination with the normalization of 
density fluctuations $\sigma_8$) provides another way to determine 
$\Odm$ \cite{bahcall1997}. One can also estimate $\Odm$ and $\Ob$ 
from the shape of the matter power spectrum. So far,
analyses of the CMB power spectra have obtained the most precise  
results on these parameters \cite{wmap9par,planck2015par}. 
Surveys like the LSST will eventually achieve similar or greater 
statistical power than analyses of the CMB do and will enable 
significant improvement over CMB-only results \cite{zhan06a,zhan06b}.

\begin{figure}
\centering
\includegraphics[width=0.425\columnwidth]{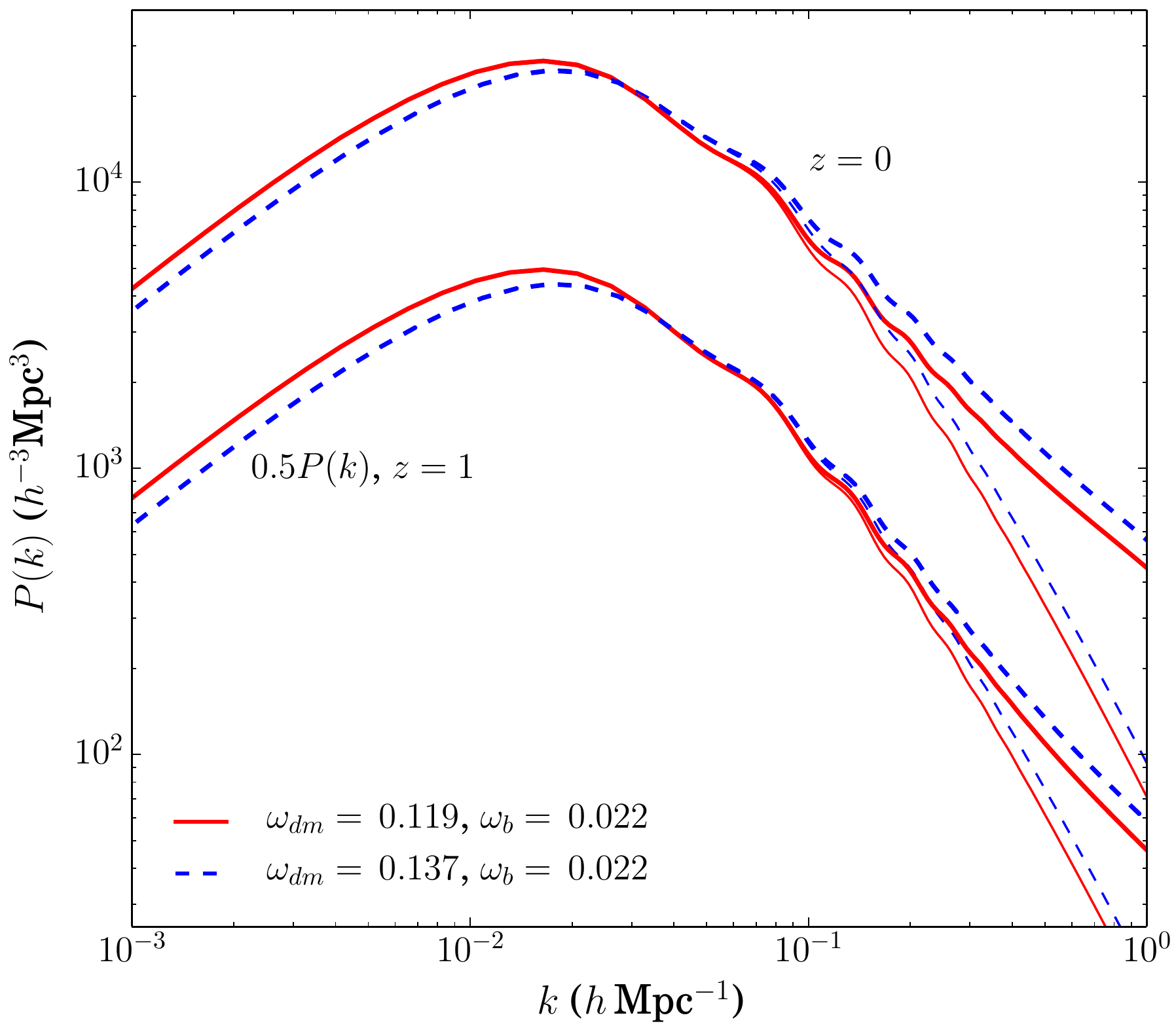}
\includegraphics[width=0.425\columnwidth]{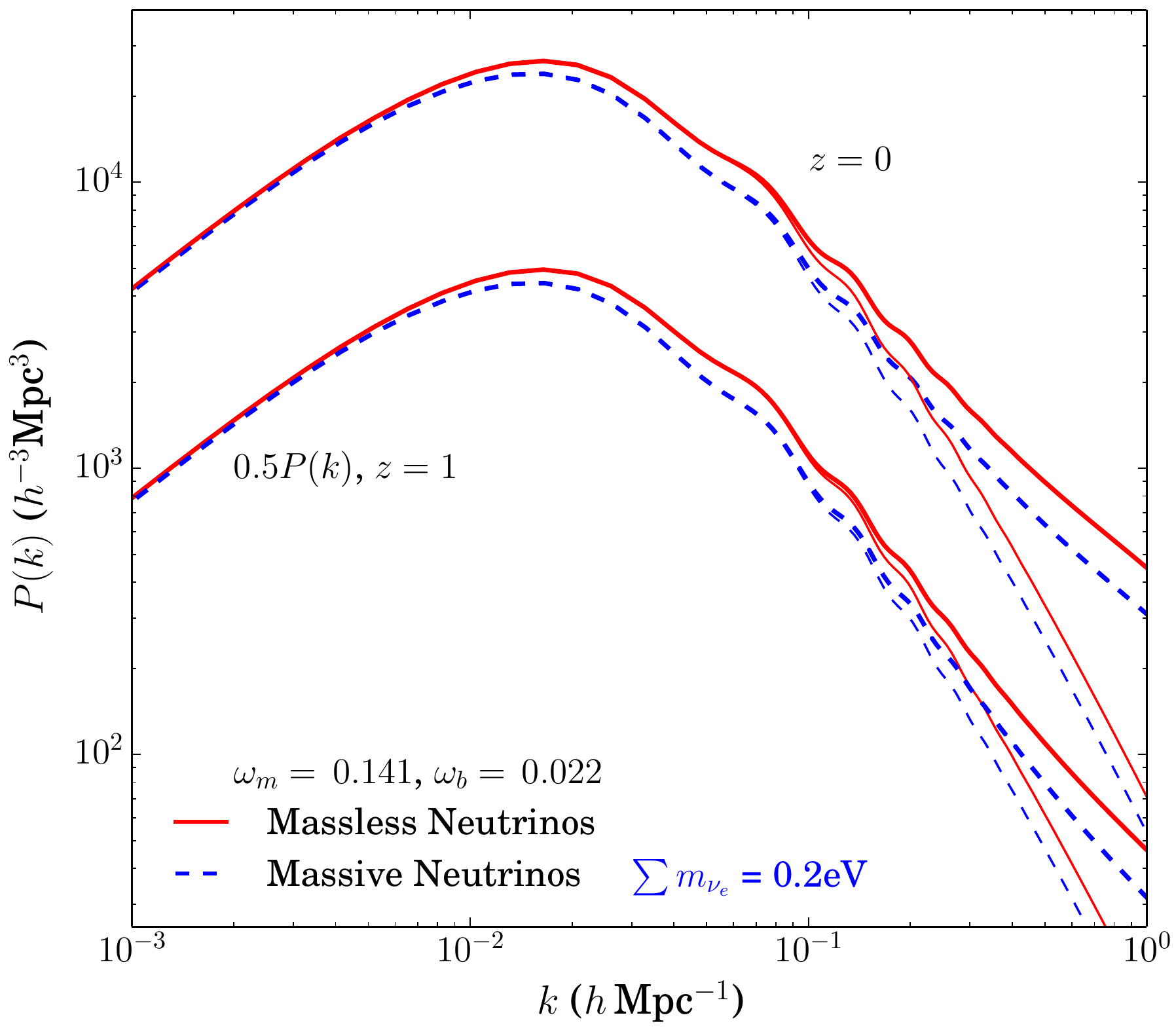}
\caption{\emph{Left}: Linear (thin lines) and nonlinear (thick lines) 
matter power spectra of two \lcdm{} models with $\odm = 0.119$ (solid 
lines) and $\odm = 0.137$ (dashed lines). \emph{Right}: Same as left,
but the $\odm = 0.137$ model is replaced by a model with massive 
neutrinos ($\sum m_{\nu_i}=0.2\,\mbox{eV}$). The two models share the 
same matter density of $\om = 0.141$. 
\label{fig:psom-mnu}}
\end{figure}

The left panel of \fref{fig:psom-mnu} illustrates the difference between 
two \lcdm{} models with $\odm = 0.119$ \cite{planck2015par} and 
$\odm = 0.137$ ($15\%$ higher than the former), 
respectively. The reduced Hubble constant takes the value of
$h=0.677$ for both models. The power spectra are calculated using
{\sc class} \cite{blas11}. Note that the shape of the matter power 
spectrum depends on $\odm$ and $\ob$ rather than $\Odm$ and 
$\Ob$ and that the different values of $\Om$ in the two models cause
a slight mismatch between their linear growth functions. 
The power spectra at $z=0$ are normalized at $k= 0.02\mpci$. 
The prominent turnover feature around 
$k_\mathrm{eq} \simeq 0.016 \hmpci$ is related to the epoch of 
matter-radiation equality ($a_\mathrm{eq} \simeq 2.8\times 10^{-4}$). 
In the radiation era, i.e., $a < a_\mathrm{eq}$, perturbations within the 
horizon were frozen. Smaller-scale perturbations entered the horizon 
earlier and experienced more suppression. After matter became dominant, 
perturbations of all scales evolved identically until nonlinearity or
non-gravitational interactions became important. A higher matter density
means that the matter-radiation equality occurred at an earlier time when 
the horizon was smaller. Therefore, the turnover scale shifts to a 
smaller scale (larger $k$) with a higher $\odm$ if $\ob$ remains
the same. In terms of parameter estimation, scales below 
the turnover generally provide stronger constraints. The wiggles around 
$k \sim 0.1 \hmpci$ in the power spectra are the BAO feature
arising from the perturbations in the same cosmic fluid that produced 
the CMB \cite{peebles70,bond84,holtzman89}. 
It is an important cosmological probe as discussed in \sref{sec:gc-bao}.

The mean density of dark matter in the universe is only one piece 
of the puzzle. More information is needed to decipher the physics of 
dark matter. Despite its success on cosmological scales, the CDM paradigm 
is at odds with observations on galactic and smaller scales, which have 
been phrased as the missing satellite problem (too many subhalos in 
simulations than observed) \cite{klypin1999,moore1999a} and the 
cusp-core problem (halos' central dark matter profile much shallower 
than predicted) \cite{moore1999b,navarro2004,gilmore2007,denaray2008}. 
There is also a related ``too big to fail'' problem (mismatch between
massive subhalos in Milky Way-like simulations and the observed bright
satellites of the Milky Way) \cite{boylan-kolchin2011}. 
Two routes to resolve the issues have been pursued: one is to 
investigate baryonic processes, such as star formation and energetic
feedback, that may lead to the observed properties of dark matter 
structures \cite{governato2012}, and the other tries to match the 
observations by replacing CDM with, for example, warm dark 
matter, self-interacting dark matter, or nonthermally produced dark
matter \cite{bode2001,spergel2000,lin2001}. Both approaches
need full development to establish a sound connection between dark 
matter theories and observations, which will be indispensable
for proper interpretations of dark matter particle 
experiments as well. Besides studies of
galaxies and their satellites, the small-scale dark matter power 
spectrum probed by quasar spectra (known as the Ly$\alpha$ forest) and 
the local dark matter density measured from the stellar distribution and 
kinematics will also provide vital information about the physics of dark matter 
\cite[e.g.,][]{bovy2012,viel2008}. 
In rare merging systems such as the Bullet cluster \cite{markevitch2002},
galaxies and dark matter may be separated from the hot X-ray gas. 
Based on the separation, kinematics, and other information, one can 
place a limit on the dark matter self-interaction cross-section 
\cite{markevitch2004, robertson2017}. The LSST and other facilities together will 
greatly expand the samples for dark matter studies and bring more 
insights with precision measurements.

\subsection{Neutrino masses}

Unlike dark energy and dark matter, neutrinos are part of the standard
model of particle physics, but the non-vanishing mass of at least one
neutrino species still needs explanation. 
Astronomical observations are crucial for determining the sum of 
neutrino masses ($\sum m_{\nu_i}$, sum over three families) 
\cite{bond1980,ma1995,wmap9par,planck2015par}. 
With neutrino oscillation results, the individual neutrino masses can be 
determined up to an ambiguity between the normal hierarchy (one species
much heavier than the other two) and the inverted hierarchy (one 
species much lighter than the other two) \cite[e.g.,][]{king2013}.
If $\sum m_{\nu_i}$ is constrained to less than $0.1\ev$, then the 
inverted hierarchy would be disfavored \cite{debernardis2009}. 
An accuracy of better than $0.02\ev$ is needed at 
$\sum m_{\nu_i} =0.06\ev$ to exclude the inverted hierarchy at more 
than 95\% confidence level \cite{hannestad2016}.

The \emph{Planck} limit on $\sum m_{\nu_i}$ with CMB, BAO, SN Ia, and 
$H_0$ data is $0.23 \ev$ ($95\%$) for a \lcdm{} universe 
\cite{planck2015par}, which is somewhat sensitive to datasets combined 
as well as model assumptions. A more recent analysis tightens the 
bound to an interesting regime of $\sim 0.1\ev$ \cite{vagnozzi2017}. 
The results indicate that the neutrinos decoupled from other matter 
before matter-radiation equality and became non-relativistic, i.e.,
matter-like, during matter domination. In this scenario, the 
matter-radiation equality is slightly delayed compared to that with 
massless neutrinos (assuming the same total matter density).
Perturbations entering the horizon 
before the transition thus have slightly less time to grow, while those
afterward are not affected. Well below their free-streaming scale, the
neutrinos do not contribute to overdensities, but they are still 
counted toward the mean matter density. Therefore, small-scale density 
perturbations continue to be suppressed after the neutrinos' 
non-relativistic transition. The overall suppression of the matter power
spectrum relative to that with massless neutrinos reaches a constant
factor of approximately $\Delta P/P \simeq -8f_\nu$ at $k \gtrsim 5 \hmpci$
for $f_\nu \lesssim 0.07$ \cite{hu1998}, where $f_\nu$ is the mean fraction 
of neutrinos in matter. The effect on intermediate scales requires detailed 
calculations \cite[e.g.,][]{lesgourgues2006}.

The right panel of \fref{fig:psom-mnu} compares the matter power spectra
of the \lcdm{} models with massless and massive neutrinos, 
respectively. For $\sum m_{\nu_i} = 0.2 \ev$, the non-relativistic 
transition would occur at $a_\mathrm{nr} = 7.5\times 10^{-3}$, corresponding 
to a scale of $k_\mathrm{nr} = 4.5\times10^{-3}\hmpci$. \Fref{fig:psom-mnu} 
indeed shows that the matter power spectra at $k \lesssim k_\mathrm{nr}$
are not affected by the neutrino masses. The shift of the turnover 
scale is barely discernible, and the suppression of the power spectra on 
small scales is pronounced. Nonlinear evolution appears to amplify the 
suppression effect of the neutrino masses. However, the nonlinear matter 
power spectra are calculated using the fitting formula from \cite{smith03}, 
which does not include massive neutrinos. Simulations are needed to 
obtain more accurate results with massive neutrinos 
\cite{ali-haimoud13,carbone2016}, and care must be taken to 
properly setup the initial conditions \cite{zennaro2017}.

Besides the sum of neutrino masses, astronomical observations also have 
sensitivity to the number of neutrino species $N_\nu$ (or the effective 
number of radiation components $N_\mathrm{eff}$). Forecasts with ideal 
assumptions predict that future surveys will reduce the uncertainties 
to roughly $0.02\ev$ on $\sum m_{\nu_i}$ and $0.05$ on $N_\mathrm{eff}$ 
\cite{song2004,debernardis2009,font-ribera14}. Therefore, it is within 
the statistical capability of these surveys to determine the sum of
neutrino masses given the minimum possible value of 
$\sum m_{\nu_i} \sim 0.06\ev$ and to positively detect extra neutrino
or radiation species. The challenge is to disentangle neutrino 
effects from those of other astrophysical and observational factors such 
as the galaxy bias and systematics.

\subsection{Primordial perturbations}

Previous subsections have shown the importance of 
the density fluctuations for determining the composition of the universe, 
though the origin of the fluctuations remains to be addressed. A generic 
prediction of inflationary models is that quantum fluctuations in the 
inflaton --- a scalar field that drove inflation --- seeded the 
density fluctuations today \cite[e.g.,][]{lyth1999}. Inflaton 
perturbations were stretched outside the horizon during inflation 
and reentered the horizon afterward as metric perturbations,
which became initial conditions of the density fluctuations. The 
primordial metric perturbations are nearly scale invariant, so 
that the corresponding initial matter power spectrum 
$P(k) \propto k^{n_\mathrm{s}}$ with the power spectral index 
$n_\mathrm{s} \sim 1$. A small departure of $n_\mathrm{s}$ from unity is 
due to  the slowly varying Hubble parameter and the inflaton field 
during inflation. Besides the inflaton perturbations, gravitational 
waves generated during inflation induce B-mode polarization in the CMB 
\cite{seljak1997,kamionkowski1997}, which have been vigorously pursued 
by a number of experiments \cite{hanson2013,polarbear2014,bicep2-2016}.

The primordial perturbations are of great interest for studies of 
inflation. Evolution of the second-order perturbations after inflation
always produces some amount of non-Gaussianity, which is quantified
by $\fnl$ --- a coefficient of the second-order term in the gravitational
potential. A null detection at the level of $\fnl \sim 1$ would rule out
the standard cosmological model \cite[e.g.,][]{bartolo2004}. As
mentioned in \sref{sec:dm}, physics within the horizon has 
altered the perturbations that reentered the horizon before the 
matter-radiation equality; only very large-scale modes are unaffected. 
Therefore surveys of huge volumes are necessary to infer the primordial 
perturbation spectrum and to probe the physics in the inflation era. 
CMB experiments have made remarkable achievements in this area 
\cite[e.g.][]{komatsu2009,planck2015ng}, though these results are based 
on a two-dimensional projection of three-dimensional modes. The LSST and other 
surveys will enable studies of the primordial perturbations with a 
comparable statistical power based on galaxy statistics in a huge 
three-dimensional volume \cite{zhan06a,alvarez2014}.

\section{Large Synoptic Survey Telescope} \label{sec:lsst}

\begin{figure}
\centering
\includegraphics[width=1.0\columnwidth]{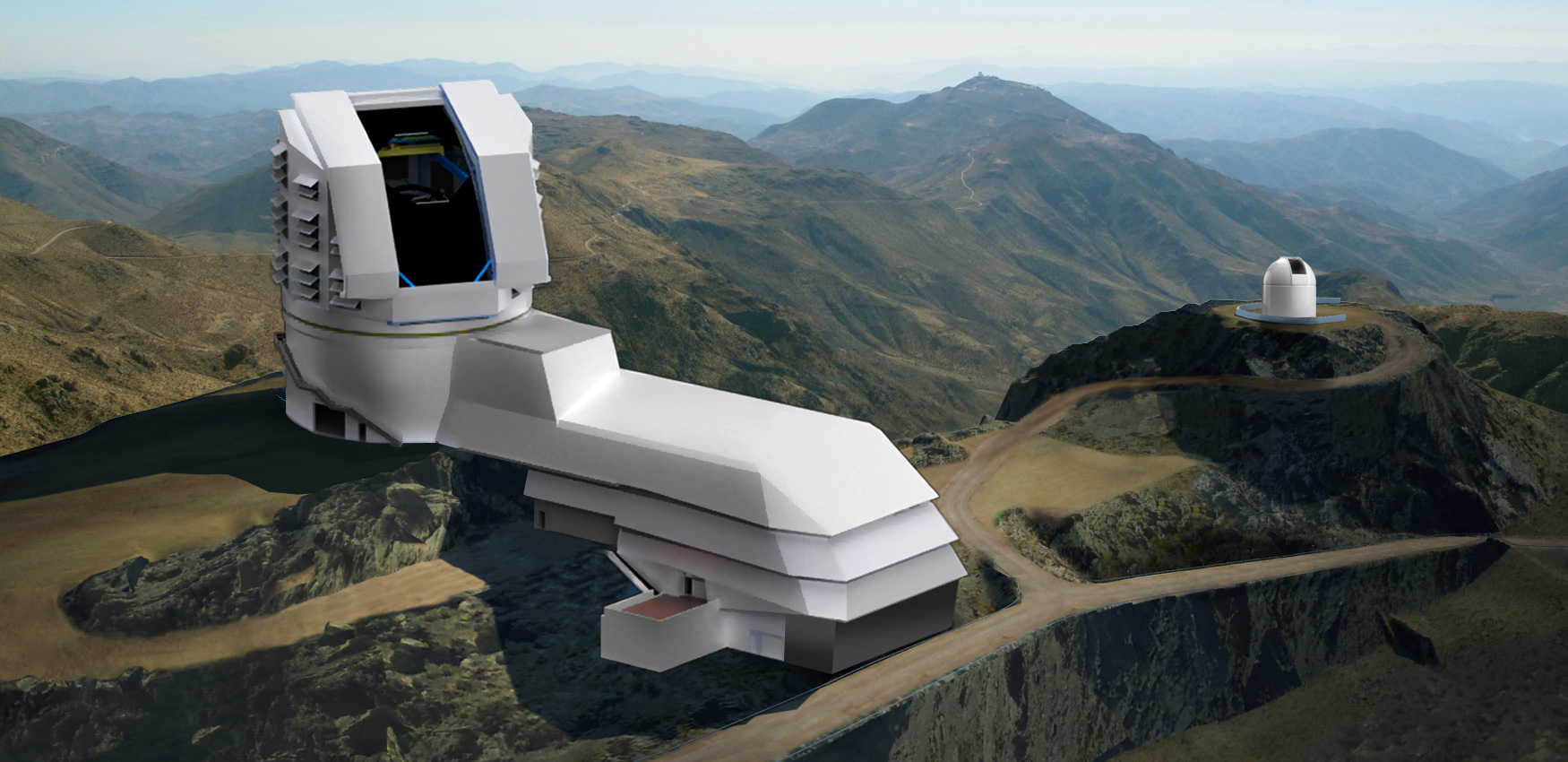}
\caption{The LSST Observatory: artist's rendering of the dome enclosure 
with the attached summit support building on Cerro Pach\'{o}n in 
Northern Chile. 
The LSST calibration telescope is shown on an adjacent rise to the right.
Image credit: the LSST Project/NSF/AURA.} 
\label{fig:observatory}
\end{figure}

The LSST is a powerful facility for cosmological studies. It
will have an $8.4\m$ ($6.5\m$ effective) primary mirror, 
a $9.6\deg^2$  FoV, and a 3.2 Gigapixel camera. An illustration of 
the LSST Observatory is shown in \fref{fig:observatory}.
This system will have unprecedented optical throughput and can image 
about 10,000$\deg^2$ of sky in three clear nights using $30\sec$
``visits'' per each sky patch twice per night, with typical $5\sigma$ depth for point sources 
of $r\sim 24.5$ (AB magnitude). The detailed cadence will be decided
by the Science Advisory Committee via simulations of observing scenarios.
The system is designed to yield high image quality as well as superb 
astrometric and photometric accuracy. 
The project is in the construction phase and will begin regular survey 
operations by 2022. The survey area will be imaged 
multiple times in six bands, \emph{ugrizy}, covering the wavelength range 
$320$--$1050\nm$. About $90\%$ of the observing time will be devoted to a 
deep-wide-fast survey mode which will uniformly observe an 18,000$\deg^2$ 
region of the southern sky over 800 times (summed over all six bands) 
during the anticipated 10 years of operations, and yield a co-added map to 
$r \sim 27.5$. These data will result in a relational database 
including 20 billion galaxies and a similar number of stars, and will serve 
the majority of the primary science programs. The remaining $10\%$ of the 
observing time will be allocated to special projects such as a very deep 
and fast time domain survey. The goal is to make LSST data products 
including the relational database of about 30 trillion observations 
of 37 billion objects available to the public and scientists around 
the world.

\subsection{Telescope and camera}

The large LSST \'etendue is achieved in a novel three-mirror design 
(modified Paul-Baker Mersenne-Schmidt system) with a very fast $f/1.2$ 
beam \cite{angel2000}. The optical design has been optimized to yield a 
large FoV, with seeing-limited image quality, across a wide wavelength 
band. Incident light is collected by an 
annular primary mirror, having an outer diameter of $8.4\m$ and inner 
diameter of $5\m$, creating an effective filled aperture of $\sim 6.5\m$ 
in diameter. The collected light is reflected to a $3.4\m$ convex secondary, 
then onto a $5\m$ concave tertiary, and finally into the three refractive 
lenses of the camera.  In broad terms, the primary-secondary mirror pair 
acts as a beam condenser, while the aspheric portions of the secondary and 
tertiary mirror act as a Schmidt camera.  The 3-element refractive optics 
of the camera correct for the chromatic aberrations induced by the 
necessity of a thick Dewar window and flatten the focal surface. All three 
mirrors will be actively supported to control wavefront distortions 
introduced by gravity and environmental stresses on the telescope. 

The LSST camera provides a 3.2 Gigapixel flat focal plane array, tiled by 
189 4K$\times$4K CCD sensors with $10\um$ pixels. 
This pixel count is a direct consequence of sampling the 
$9.6 \deg^2$ FoV ($0.64\m$ diameter) with $0.2\arcsec\times0.2\arcsec$
pixels (Nyquist sampling in the best expected seeing of $\sim 0.4\arcsec$). 
The sensors are deep-depleted high-resistivity silicon back-illuminated 
devices with a highly segmented architecture that enables the entire array 
to be read in 2 seconds. The CCDs are grouped into 3$\times$3 rafts, 
each containing its own dedicated electronics. The rafts are mounted on a 
silicon carbide grid inside a vacuum cryostat, with an intricate thermal 
control system that maintains the CCDs at an operating temperature of 
$173\K$. The entrance window to the cryostat is the third of the three 
refractive lenses in the camera. The other two lenses are mounted in an 
optics structure at the front of the camera body, which also contains a 
mechanical shutter, and a carousel assembly that holds 
five large optical filters. The sixth optical filter can 
replace any of the five via a procedure accomplished during daylight hours. 

\begin{figure}
\centering
\includegraphics[width=1.0\columnwidth]{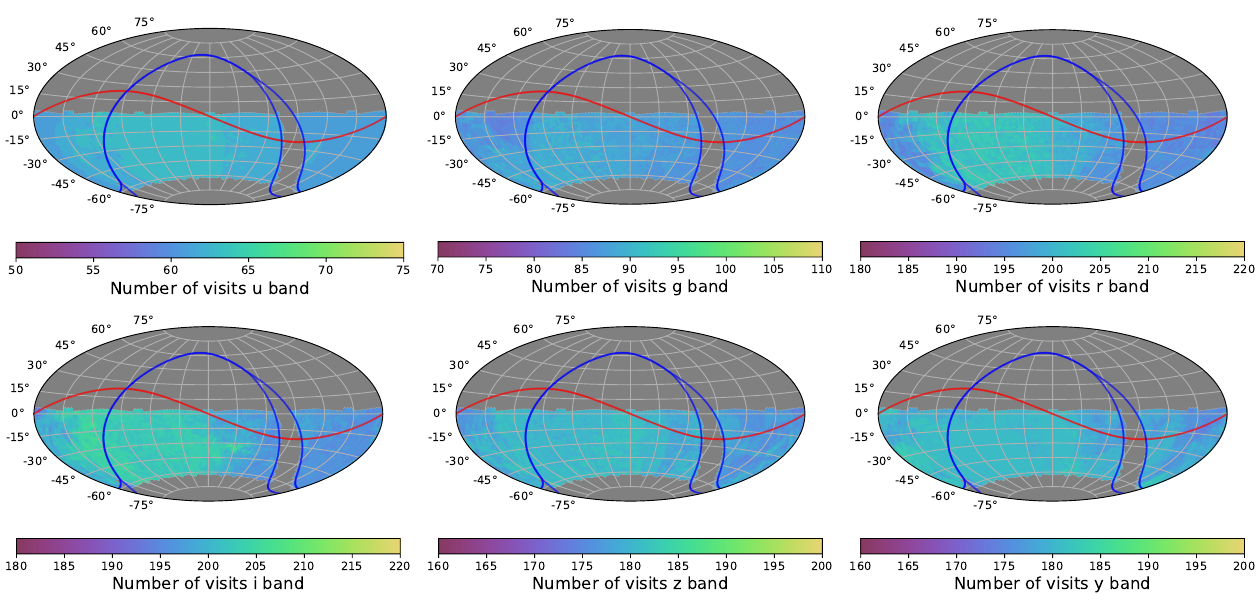}
\caption{The distribution of the 6 band visits on the sky for a simulated 
realization of the baseline cadence. 
Image credit: Lynne Jones and the LSST Project/NSF/AURA.}
\label{fig:6bandVisitsSky}
\end{figure}

\subsection{Survey plan and performance}

The main deep-wide-fast survey (typical single visit depth of $r\sim24.5$)
will use about 90\% of the observing time. The remaining 10\% of the 
observing time will be used to obtain improved coverage of parameter space 
such as very deep observations. These deeper fields (deep-drilling fields)
will aid in statistical completeness studies for the main survey, since 
they will likely have deep spectroscopy as well as infrared coverage from 
other facilities. 
The observing strategy for the main survey will be optimized for 
homogeneity of depth and number of visits. In times of good seeing and at 
low airmass, preference will be given to $r$-band and $i$-band observations 
which are used in WL.  
The visits to each field will be widely distributed in position angle on 
the sky and rotation angle of the camera in order to minimize systematic 
effects on the point-spread function (PSF), which could introduce shear 
systematics in faint galaxies. Simulations of LSST operations use actual 
weather data from the Chilean site. The detailed cadence in time and space 
across the sky is being optimized with these simulations. We show one 
such simulation of the 6 band coverage in \fref{fig:6bandVisitsSky}
\cite{connolly2014,jones2014}.

The universal cadence proposal excludes observations in a region of 
1000$\deg^2$ around the Galactic Center, where the high stellar density 
leads to a confusion limit at much brighter magnitudes than those 
attained in the rest of the survey.  The anticipated total number of 
visits for a ten-year LSST survey is about 2.8 million. The per-band allocation of these visits 
is shown in \tref{tab:lsst_baseline}. The adopted time allocation (see 
\tref{tab:lsst_baseline}) includes a slight preference for the $r$ and 
$i$ bands because of their dominant role in star/galaxy separation and 
WL measurements. 

\begin{table} 
\caption{The LSST Baseline Design and Survey Parameters\label{tab:lsst_baseline}}
\begin{tabular}{|l|l|}
\hline  
   Quantity                         &     Baseline Design Specification    \\
\hline  
Optical Config.                           &  3-mirror modified Paul-Baker        \\
Mount Config.                            &  Alt-azimuth          \\
Final f-ratio, aperture                 &  f/1.234, $8.4\m$                \\
FoV, \'etendue              &  $9.6 \deg^2$,   319 m$^2$deg$^2$     \\
Plate Scale                                  &  $50.9 \um/\mbox{arcsec}$ (0.2'' pix)  \\
Pixel count                                  &  3.2 Gigapix  \\
Wavelength Coverage                   &  320 -- 1050 nm, $ugrizy$             \\
Single visit depths, design $^a$  &  23.9, 25.0, 24.7, 24.0, 23.3, 22.1    \\
Mean number of visits$^b$          &  56, 80, 184, 184, 160, 160               \\ 
Final (coadded) depths$^c$         &  26.1, 27.4, 27.5, 26.8, 26.1, 24.9     \\
\hline                         
\end{tabular}
\\ \vskip 0.05in
$^a$ Design specification from the Science Requirements Document (SRD) 
\cite{lsst_srd} for $5\sigma$ depths for point sources in the $ugrizy$ bands,
respectively. The listed values are expressed on AB magnitude scale, and 
correspond to point sources and fiducial zenith observations (about 0.2 mag 
loss of depth is expected for realistic airmass distributions). \\
$^b$ An illustration of the distribution of the number of visits as a 
function of bandpass, taken from Table 24 in the SRD.  \\
$^c$ Idealized depth of coadded images, based on design specification for 
5$\sigma$ depth and the number of visits in the penultimate row (taken from 
Table 24 in the SRD). 
\vskip 0.2in          
\end{table}

Precise determination of the PSF across each image, 
accurate photometric and astrometric calibration, and continuous monitoring 
of system performance and observing conditions will be needed to reach the 
full potential of the LSST mission. The dark energy science requires 
accurate photometric redshifts, so the LSST photometry will be calibrated 
to unprecedented precision. 
Auxiliary instrumentation, including a $1.5\m$ calibration telescope, 
will provide the calibration parameters needed for image 
processing, to calibrate the instrumental response of the LSST hardware
\cite{stubbs2006}, and to measure the atmospheric optical depth as 
a function of wavelength along the LSST line of sight
\cite{stubbs2007}.

\subsection{Data products}

The rapid cadence and length of the LSST observing program will produce
approximately 15 TB per night of raw imaging data.
The large data volume, the time domain aspects, and the complexity of processing involved makes
 it impractical to rely on the end users for the data reduction. Instead, the data collected by the 
LSST system will be automatically reduced to scientifically useful catalogs and images.
Over the ten years of LSST operations and 11 data releases, this processing will result in 
cumulative \emph{processed} data of about 500 PB for imaging, and over 50 PB for the
catalog databases. The final data release catalog database alone is expected 
to be approximately 15 PB in size.

Data collected by the LSST telescope and camera will be automatically 
processed to \emph{data products} --- catalogs, alerts,
and reduced images. These products are designed to
enable a large majority of LSST science cases, without the need to
work directly with the raw pixels.  We give a high-level overview of
the LSST data products here; further details may be found in the LSST
Data Products Definition Document \cite{lsst_dpdd}, 
which is periodically updated. These data will be served via a relational database.

Two major categories of data products will be produced and delivered by 
the LSST: Level 1 and Level 2.   Level 1 are time domain: data products which support the 
discovery, characterization, and rapid follow-up of time-dependent 
phenomena. Level 2 data products are most relevant to cosmology:  
they are designed to enable systematics- and flux-limited 
science, and will be made available in annual Data Releases. 
These will include the single-epoch images, deep coadds 
of the observed sky, catalogs of objects detected in the LSST data, catalogs 
of sources (the detections and measurements of objects on individual 
visits), and catalogs of ``forced sources" -- measurements of flux on 
individual visits at locations where objects were detected by the LSST or 
other surveys. LSST Level 2 processing will 
rely on multi-epoch model fitting, or \emph{MultiFit}, to
perform near-optimal characterization of object properties. 
Although the coadded images will be used to perform object \emph{detection}, the
\emph{measurement} of their properties will be performed by
simultaneously fitting (PSF-convolved) models to single-epoch
observations. An extended source model --- a constrained linear
combination of two S\'ersic profiles --- and a point source model with
proper motion --- will generally be fitted to each detected object.

For the extended source model fits, the LSST will
characterize and store the shape of the associated likelihood surface
(and the posterior) -- not just the maximum likelihood values and
covariances. The characterization will be done by sampling,
with up to $\sim200$ (independent) likelihood samples retained for
each object. For reasons of storage cost, these samples
may be retained only for those bands of greatest interest for
WL studies.

While a large majority of science cases will be adequately served by
Level 1 and 2 data products, a limited number of highly specialized
investigations may require custom, user-driven, processing of LSST
data. This processing will be most efficiently performed at the 
LSST Archive Center, given the size of the LSST data set and the
associated storage and computational challenges. To enable such use
cases, the LSST DM system will devote the equivalent of 10\% of its
processing and storage capabilities to creation, use, and federation
of so-called ``Level 3'' (user-created) data products. It will also 
allow the science teams to use the LSST database infrastructure to store 
and share their results. 
The LSST Archive Center and U.S. data access center will be at the 
National Center for Supercomputing Applications (NCSA) \cite{ncsa_lsst}.
Users will access the LSST data through a Data Access Center web portal, 
a Jupyter Notebook interface, and machine accessible web application 
programming interfaces. The web portal will provide data access and
visualization services, and the Notebook 
interface will enable more sophisticated data analysis.
  
\subsection{Engaging the community}

The LSST database and the associated object catalogs will be made available 
to the U.S. and Chilean scientific 
communities and to international partners with no proprietary period. 
The LSST project has been working with international partners 
to make LSST data products available worldwide. User-friendly tools for data 
access and exploration will be provided by the LSST data management 
system.  This will support user-initiated queries and will run on LSST 
computers at the archive facility and the data access centers. 

Because of the volume of the LSST data, statistical noise will reach 
unprecedented low levels so that some investigations will be limited by
systematics, although at a level far below those of previous surveys. 
Thus, those 
investigations will require organized teams working together to optimize 
science analyses. LSST science collaborations have been established in core 
science areas. The LSST DESC includes members with interests in dark energy 
and related topics in fundamental physics.  

The LSST Project is actively seeking and implementing
input by the LSST science community.  The LSST science collaborations
in particular have helped develop the LSST science case and continue
to provide advice on how to optimize their science with choices in
cadence, software,  and data systems. During the commissioning period, the Science 
Collaborations will play a role in the system optimization. 
The LSST Science Advisory Committee provides a 
formal dialogue with the science community. 
This committee also deals with technical topics of interest to 
both the science community and to the LSST Project, and shares responsibility for 
policy questions with the Project Science Team.

\section{LSST Probes of Cosmology} \label{sec:probes}

With its deep-wide-fast multiband imaging survey, the LSST will 
enable multiple probes simultaneously for cosmological studies. 
Here we discuss several probes that have been studied extensively: 
WL, LSS (or BAO\footnote{Much of the cosmological constraining power 
of the LSS in the LSST \phz{} galaxy sample is from the BAOs in the 
galaxy angular power spectra, so we use LSS and BAO interchangeably 
for the LSST unless it is necessary to make a distinction.}), 
SNe Ia, galaxy clusters, and strong lensing. 
WL is considered the most powerful among these
probes, and, at same time, it also imposes the most stringent requirements 
on the project (the telescope, data management, and operations) and 
beyond (e.g., analysis pipelines, computing infrastructure, simulations,
theories, and so on). 
Since dark energy, or cosmic acceleration in general, has become a common 
science driver of almost every large extragalactic survey, these probes are 
often associated with dark energy, even though they are also 
sensitive to various elements of cosmology as can be inferred from 
\sref{sec:frontiers}.

It is worth emphasizing the strength of utilizing multiple probes of the
same survey as well as those of different surveys to address the fundamental 
questions about the universe. These probes will not only form interlocking
cross-checks but also provide means of calibrating mutual systematics, 
which would be a common challenge for surveys like the LSST.
The combination of WL and BAO is a particularly 
effective probe which breaks degeneracies and suppresses systematics.

We note that one probe can be analyzed using the 
methods of another probe. For example, in addition to their mass function,
galaxy clusters may be studied for the LSS that they trace and their 
WL and strong lensing effects. 
Different statistics can also be applied to the same 
observables of each probe. For brevity, we introduce below only the 
most discussed method(s) and statistic(s) for each probe.

\subsection{Large-scale structure (BAO)}
\label{sec:gc-bao}

Statistical analysis of the large-scale galaxy distribution is the 
main tool of LSS studies. Although LSS based on imaging data alone 
does not place strong constraints on the dark energy EoS, 
it is generally more sensitive than other LSST probes to cosmological 
parameters that affect the shape of the underlying matter power spectrum. 
LSS is a quite mature field with decades of studies 
(e.g., \cite{peebles80}), and its formulism can find 
applications in WL as well. 

The LSST ``gold sample'' is expected to contain at least 2.6 billion galaxies 
at $i_\mathrm{AB} \le 25.3$ mag. The redshift distribution is expected 
to follow $n(z) \propto z^2\exp(-3.2z)$ with an integrated surface number
density of more than $40 \arcmin^{-2}$ \cite[section 3.7]{lsst_sb}. 
Although it is advantageous to utilize the full posterior probability 
distribution of the photometric redshift (\phz) of each galaxy 
\cite{myers09,abrahamse11,sheldon12}, we assume for convenience that a 
\phz{} is assigned to each galaxy. The science 
requirement on the \phz{} rms error per galaxy is $\sigma_z(z) \le \sigma_{z0}(1+z)$ 
with $\sigma_{z0}=0.05$, and the goal is to achieve 
$\sigma_z(z) \sim 0.02(1+z)$. However, even with $\sigma_{z0}$
as small as $0.02$, the line-of-sight clustering 
information is still severely suppressed at $k \gtrsim 0.02\hmpci$ 
\cite{zhan06a,zhan08}. Therefore, it is more practical to focus on 
angular clustering of the galaxies between \phz{} bins.

With the Limber approximation \cite{limber54}, the galaxy angular power 
spectrum $C_{ij}(\ell)$ is given by
\begin{eqnarray} \label{eq:gaps}
C_{ij}(\ell) &=& \frac{2\pi^2}{c\ell^3} \int_0^\infty \mathrm{d} z\, H(z) 
\Da(z) W_i(z) W_j(z) \Delta^2(k; z)  + 
\delta_{ij}^\mathrm{K} \frac{1}{\bar{n}_i} \\ \label{eq:Wg}
W_i(z) &=& b(z) n_i(z)/\bar{n}_i, 
\end{eqnarray}
where $i$ and $j$ identify the \phz{} bins, $\ell$ is the multipole 
number, $k = \ell/\Da$, $\delta_{ij}^\mathrm{K}$ is the Kronecker delta 
function, $b(z)$ is the linear galaxy clustering bias, $n_i(z)$ is the 
redshift distribution of galaxies in bin $i$, and 
$\bar{n}_i=\int n_i(z) \rmd z $. The last term on the right-hand side
of \eref{eq:gaps} is the 
shot noise due to discrete sampling of the continuous density field with 
galaxies. The covariance between the power spectra $C_{ij}(\ell)$ 
and $C_{mn}(\ell)$ per angular mode is given by
\begin{equation} \label{eq:covgps}
\mathrm{Cov}\left[C_{ij}(\ell), C_{mn}(\ell)\right] = 
C_{im}(\ell) C_{jn}(\ell) + C_{in}(\ell) C_{jm}(\ell),
\end{equation}
and the rms error of $C_{ij}(\ell)$ is approximately
\begin{equation} \label{eq:lss:gpserr}
\sigma[C_{ij}(\ell)] = \left[\frac{C_{ii}(\ell)C_{jj}(\ell) + 
C_{ij}^2(\ell)}{f_\mathrm{sky}(2\ell + 1)}\right]^{1/2},
\end{equation}
where $f_\mathrm{sky}$ is the fraction of sky covered by the survey, e.g.,
$f_\mathrm{sky}=0.44$ for the LSST.

The scale dependence of the galaxy bias becomes more pronounced 
below tens of $\himpc$, so one has to limit the application of 
\eref{eq:gaps} to large scales, or model the bias in detail 
\cite{hu04a,zheng07,cacciato12}, or determine the galaxy bias with 
higher-order statistics \cite{fry94,verde02}.
\Eref{eq:gaps} is also inaccurate on very large scales because 
the Limber approximation breaks down; if primordial non-Gaussianity is 
considered, the galaxy bias needs to be replaced by an effective bias 
that scales roughly as $k^{-2}$ on very large scales 
\cite{dalal2008,matarrese08}.

\begin{figure}
\centering
\includegraphics[width=0.9\columnwidth]{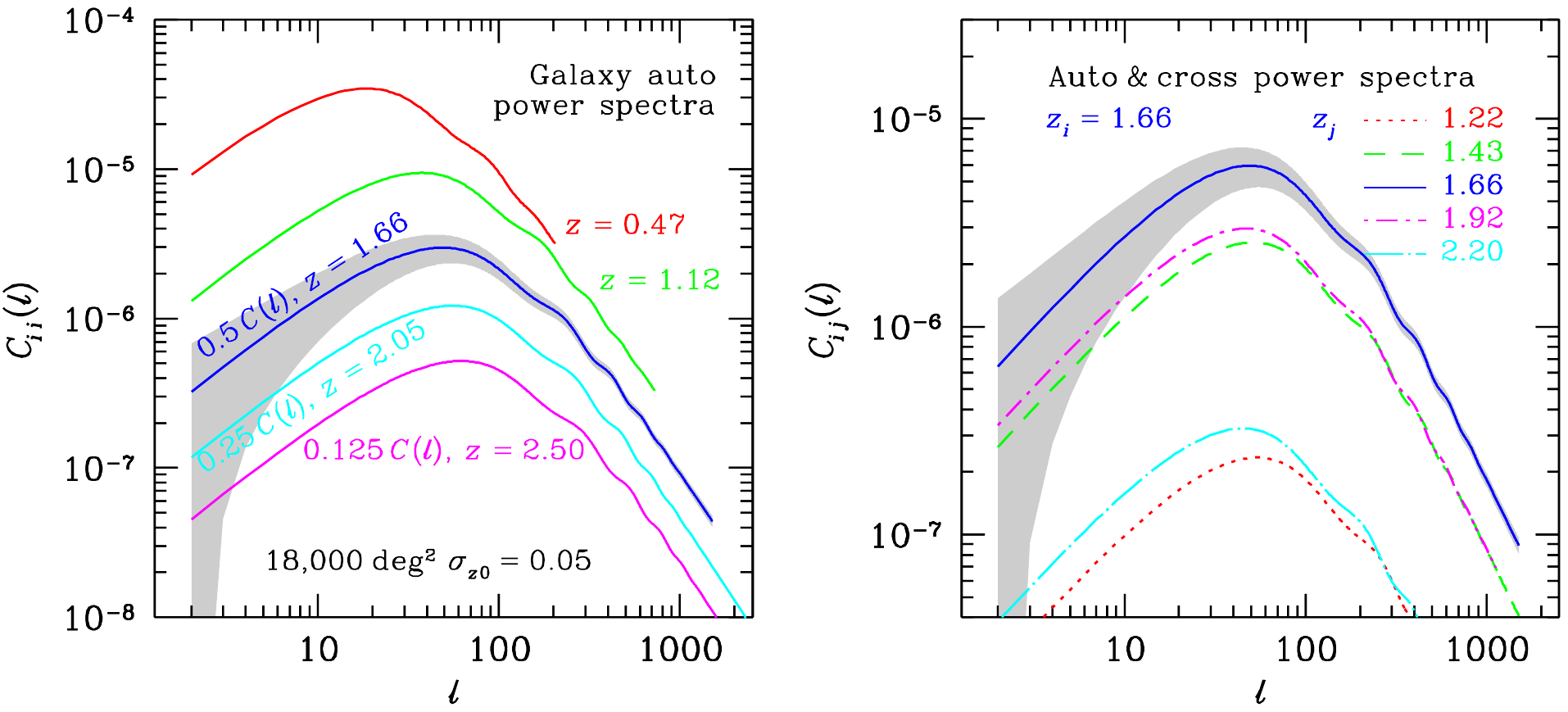}
\caption{{\it Left panel}: Galaxy angular auto power spectra in five
redshift bins (shifted for clarity). The central \phz{} of each bin 
is as labeled. The gray area indicates the statistical error 
(cosmic variance and shot noise) \emph{per multipole} for the bin 
centered at $z = 1.66$. 
{\it Right panel}: Cross power spectra between bin 
$i$ centered at $z = 1.66$ and bin $j$ centered at 
$z = 1.22$ (4th neighbor, dotted line), $1.43$ (2nd 
neighbor, dashed line),  $1.92$ (2nd neighbor, dash-dotted line), 
and $2.20$ (4th neighbor, long-dash-dotted line). The auto power 
spectrum at $z=1.66$ is the same as that in the left panel. 
Figure adapted from \cite{lsst_sb} with updated survey 
data model.
\label{fig:bcl}}
\end{figure}

As an example, we show several galaxy angular power spectra in 
\fref{fig:bcl}. The LSST gold sample is assigned to 30 bins from \phz{} 
of $0.15$ to $3.5$ with the bin width proportional to $1+z$ in order to 
match the \phz{} rms error. Five auto power spectra in 
their respective bins are given in the left panel of \fref{fig:bcl}.
The power spectra are truncated at the high-$\ell$ end where nonlinear
evolution starts to become important. Details of the calculations including
the \phz{} treatment can be found in \cite{zhan06b}.
The right panel of \fref{fig:bcl} shows four cross power spectra 
between the bin centered on $z = 1.66$ and its neighbors, and the auto 
spectrum at $z = 1.66$ is included for reference.
The amplitude of the cross power spectrum is largely determined by the overlap
between the two bins in true redshift space, so it decreases rapidly
with the bin separation under the Gaussian \phz{} model. This property 
can help calibrate  the \phz{} error distribution 
\cite{zhan06b,schneider2006}.

One can clearly identify the BAO wiggles in the galaxy angular power spectra 
in \fref{fig:bcl} despite the projection of three-dimensional fluctuations 
onto the sphere. These wiggles are an imprint of acoustic waves in the tightly 
coupled cosmic fluid before the universe became sufficiently cool to form 
neutral hydrogen around $z \simeq 1100$. The primary CMB temperature 
anisotropy is a snapshot of these acoustic waves at the last scattering 
surface, which is characterized by the sound horizon $r_s \sim 150\,\mpc$ 
at that time \cite{peebles70,bond84,holtzman89}. 
The BAO feature is mainly a function of the matter density and the 
baryon density, and its scale is large enough to remain nearly unchanged
in comoving space since $z \simeq 1100$. Therefore, it can serve as a
standard ruler to measure angular diameter distances (and Hubble parameters
as well if accurate redshifts can be obtained) and to further constrain 
cosmological parameters 
\cite{eisenstein98b,cooray01b,blake03,hu03b,linder03b,seo03}. Recent
spectroscopic BAO surveys have measured several distances up 
to $z=2.3$ with precision of a few percent
\cite{beutler11,font-ribera14b,anderson14,delubac15}. 
The LSST gold sample will enable more distance measurements at the 
percent level up to $z = 3$ with BAOs only \cite{zhan2009a}.

\begin{figure}
\centering
\includegraphics[width=0.6\columnwidth]{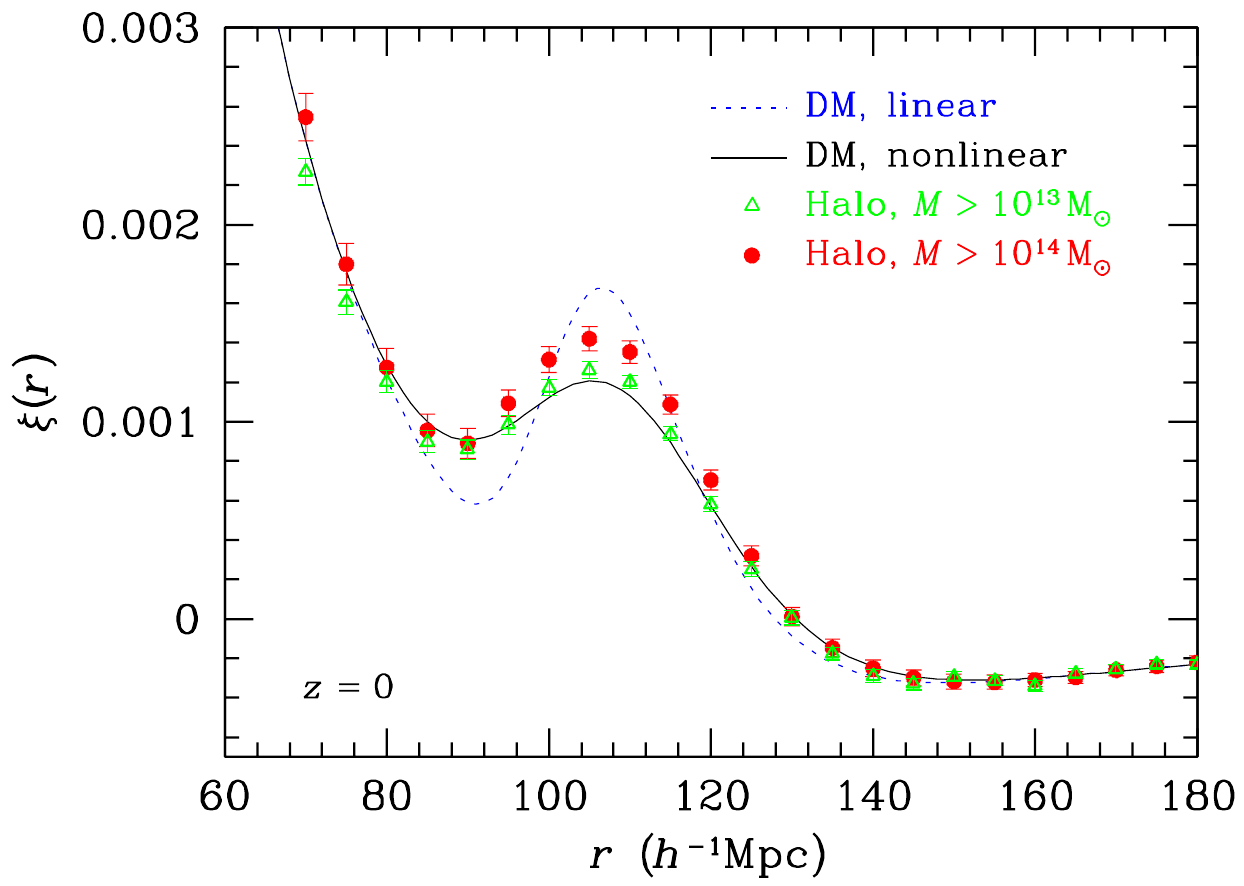}
\caption{Damping of the BAO signal at $z=0$. The dashed line and the solid 
line represent, respectively, the linear dark matter correlation function
and the nonlinear dark matter correlation function. The correlation function 
of $M > 10^{14}\Msun$ halos from \emph{N}-body simulations is plotted with
filled circles, and that of $M > 10^{13}\Msun$ halos is plotted with open 
triangles. These correlation functions are normalized in the range of 
$160$--$180\mpc$. Data from \cite{wang13}.
\label{fig:halobao}}
\end{figure}

It is recognized that nonlinear evolution modifies the BAO feature 
over time by damping its amplitude and shifting its scale by a fraction of
a percent
\cite{seo2008,seo2010,sherwin2012,mccullagh2013}. These effects can be
calibrated with simulations. \Fref{fig:halobao} demonstrates the damping 
effect on the correlation functions at $z=0$. The wiggles in the power 
spectrum become a peak in the correlation function near $r_s \sim 110\himpc$. 
The BAO signal in the nonlinear dark matter correlation function (solid
line, calculated with renormalized perturbation theory \cite{crocce2008})
has less contrast than that in the linear dark matter correlation 
function (dotted line), and the shift of the BAO scale is not easily
discernible. Although not shown, the correlation 
function of dark matter particles in \emph{N}-body simulations agrees well 
with the predicted nonlinear dark matter correlation function. 
The BAO signal of $M>10^{14}\Msun$
halos (solid circles), however, experiences less damping. It is
interesting to note that current galaxy BAO measurements all show stronger 
BAO signal than expected (e.g., 
\cite{beutler11,font-ribera14b,anderson14,delubac15}). 
Such an effect has little impact on cosmological constraints if one only 
uses the BAO scale information. However, a proper analysis utilizing the 
full correlation function(s) could gain significantly over \emph{Planck}
on parameters such as $\om$, $\ob$, and $\Ok$ \cite{zhan06b}.

\subsection{Weak gravitational lensing (WL)}
\label{sec:wl}

Gravitational lensing is a ubiquitous phenomenon. In most 
cases, the distortion on background sources, e.g., galaxies, is so weak 
that it can only be detected statistically
\cite{wittman2000,vanwaerbeke2000,bacon2000}. Since the lensing effect is 
caused by a foreground mass distribution, its statistics reflect those 
of the foreground and can be used to probe cosmology. 
In this subsection, we only consider two-point 
statistics of the shear and magnification effects of WL. 
Interested readers are referred to \cite{bartelmann01,munshi08} for 
a wealth of WL topics. 

Given a lensed image in coordinates $\boldsymbol{\theta}$, one can 
recover the unlensed source in coordinates $\boldsymbol{\beta}$ with 
the local distortion matrix
\beq \label{eq:lens-jacobian}
\bit{A} = \frac{\partial \boldsymbol{\beta}}{\partial \boldsymbol{\theta}}
= \left(\begin{array}{cc} 1 - \kappa - \gamma_1 & -\gamma_2 \\
-\gamma_2 & 1 - \kappa + \gamma_1 \end{array} \right),
\eeq
where the convergence $\kappa$ and shear components $\gamma_i$ are 
related to the two-dimensional lensing potential 
$\psi(\boldsymbol{\theta})$ via
$\kappa = \frac{1}{2}\left(\partial_{11}^2\psi+\partial_{22}^2\psi\right)$, 
$\gamma_1 = \frac{1}{2}\left(\partial_{11}^2\psi-\partial_{22}^2\psi\right)$, 
and $\gamma_2 = \partial_{12}^2\psi$. The convergence for a source at 
$(\boldsymbol{\theta},z)$ is a weighted integral of the overdensity 
along the line of sight
\beq
\kappa(\boldsymbol{\theta},z) = \frac{3H_0^2\Om}{2c\Da(z)}
\int_0^{z}\rmd z'\,\frac{\Da(z',z)\Da(z')}{H(z')}
\delta(\boldsymbol{\theta},z') (1+z').
\eeq
The lensing magnification is given by 
$\mu = \left[(1-\kappa)^2-\gamma^2\right]^{-1}$. In the WL regime, 
$|\kappa|$ and $|\gamma|$ are much smaller than unity, 
so that $\mu \simeq 1+2\kappa$. For convenience, we use the word
``magnification'' to also cover the case of demagnification.

Assuming that galaxies are randomly oriented in the absence of lensing, 
one can estimate the shear from the average galaxy ellipticities within 
an area of appropriate size. In practice, shear measurement is rather 
difficult, and it constitutes a major source of systematic errors for 
WL. Much effort is being made to improve the accuracy of shear 
measurement \cite[e.g.,][]{great3,schneider2015,bernstein2016,sheldon2017}. 
Magnification changes the angular size of the background 
in different directions while conserving the $4\pi$ solid angle 
as viewed by the (freefall) observer. To be specific, let $\mu > 1$ 
in a particular direction. On the one hand, background galaxies are 
enlarged in this direction without altering their surface brightness, 
so they appear brighter in terms of total flux. 
Since the apparent magnitude is usually important for sample selection, 
faint galaxies that do not actually meet the magnitude criterion could 
make the cut with the slight lensing boost. This effect would increase 
the galaxy overdensity in the direction. On the other hand, magnification 
also increases the angular separation between the background galaxies, 
reducing their apparent surface density. These two competing effects 
do not cancel in general, and the resulting galaxy number density 
fluctuations of the sample is another useful probe of cosmology
\cite{moessner1998,jain2002,zhang2006,vanwaerbeke2010}.

Angular power spectra (or two-point correlation functions) of 
shear signals are the primary statistics of WL. They
can be written in a similar form as \eref{eq:gaps}
\bea \label{eq:saps}
C_{ij}^{\gamma\gamma}(\ell) &=& \frac{2\pi^2}{c\ell^3} \int_0^\infty 
\! \rmd z\, H(z) D_\mathrm{A}(z) W_i^\gamma(z) W_j^\gamma(z) \Delta^2(k; z) + 
\delta_{ij}^\mathrm{K}\frac{\sigma_\gamma^2}{\bar{n}_i},  \\  \label{eq:Ws}
W_i^\gamma(z) &=& \frac{3}{2}\frac{\Omega_\mathrm{m}H_0^2}{H(z)}
\frac{D_\mathrm{A}(z)}{a\,c} \int_z^\infty \! \rmd z'\, 
\frac{n_i(z')}{\bar{n}_i}
\frac{D_\mathrm{A}(z,z')}{D_\mathrm{A}(z')}, 
\eea
where $\sigma_\gamma \sim 0.2$ is the rms shear of galaxies' shape. 
The last term in \eref{eq:saps} is known as the shape noise, analogous 
to the shot noise in galaxy power spectra. The window function 
$W_i^\gamma(z)$ is a broad function in redshift, peaking roughly midway 
between the sources and the observer, so the cross power spectrum between 
two \phz{} bins has an amplitude close to that of the auto power spectrum
in the lower \phz{} bin.  

It is customary to decompose the effects of shear systematics on the 
power spectrum into multiplicative and additive components. The former
arises from errors that are correlated with the true shear 
signal, whereas the latter is uncorrelated with the signal. 
The observed shear power spectra are then given by
\beq  \label{eq:aps-sys}
\tilde{C}_{ij}^{\gamma\gamma}(\ell) = (1+f_i)(1+f_j)C_{ij}^{\gamma\gamma}(\ell) 
+ C_{ij}^\mathrm{add}(\ell),
\eeq
where $f_i$ and $f_j$ account for the multiplicative errors, and 
$C_{ij}^\mathrm{add}(\ell)$ is the power spectrum of the additive errors. 
In principle, these forms of errors also exist in the galaxy power 
spectra, but results from existing galaxy surveys suggest that they 
are  sub-dominant. For the LSST, the multiplicative
errors of the shear power spectra need to be controlled below 
$0.004$ \cite{huterer06, massey2013}, which is challenging. The dominant issue 
for ground-based imaging is the PSF correction and its misestimation.
This is particularly important 
for tomographic cosmic shear where the evolution of the multiplicative correction
to the observed shear must correctly track the evolving size distribution
of the source galaxies with redshift. This low level of correction can be 
achieved for a deep tomographic WL cosmic shear survey via simulations of 
the observations \cite{jee2013, jee2016}.  For additive shear it is not trivial to 
accurately determine the scale dependence of $C_{ij}^\mathrm{add}(\ell)$, 
but the level of additive errors estimated from image simulations 
\cite{chang13} suggests that it would not be overly important for 
the LSST \cite{petri14}. There are also small additive and multiplicative PSF errors
introduced by charge transport anomalies in the detector, however these 
can be well calibrated and removed to first order in pixel processing.
These errors, particularly the multiplicative
systematic, can be suppressed via observing strategy.  The amplitudes of these
errors decrease rapidly in good image quality: this  translates to low airmass
observing  and a strategy of reimaging areas with previous poor seeing exposures.
Special dithering of exposures in subsequent re-visits to a field, in 
position and camera angle, are planned.

The fact that WL measures directly the gravitational effect of all the 
matter and that it is sensitive to both the expansion of the universe and 
the growth of cosmic structures make it a powerful probe for the LSST, 
though it also has a number of challenges besides shear 
estimation. A major source of error is uncertainties in the 
\phz{} error distribution or, equivalently, those in the true redshift
distribution of galaxies in each \phz{} bin.
Because the lensing kernel is very broad, a small error in $n_i$, e.g.,
a shift of its central redshift by $\Delta z\sim 0.01$, has practically
no effect on the shear power spectra. However, the cosmological 
interpretation would be sufficiently different even with such a small
redshift error that dark energy constraints would be degraded 
considerably. It is recognized that the uncertainties in the mean 
and the width of the true-redshift distribution of galaxies in each 
\phz{} bin should not exceed a few $10^{-3} (1+z)$ for the LSST 
\cite{ma06,huterer06,zhan06b}. To achieve this goal, calibrations of 
both individual galaxy redshift \cite{bernstein2010,hearin2010} and the
overall redshift distribution of the galaxies in each \phz{} bin are 
needed. The latter can be done to the required precision with correlations 
between spatially overlapping \phz{} and spectroscopic samples of brighter 
galaxies over small patches of the survey area 
\cite{newman2008,matthews2012}. Correlations between different \phz{} 
bins arising from incorrect assignment of the galaxies will also be 
helpful for the calibration \cite{zhan06b,schneider2006,zhang2010}. 

At the surface density of galaxies in the LSST gold sample, a small 
percentage will be partially
blended with other galaxies in the sample.  This can cause outliers in
photo-z and generate about a 7\% increase in WL shear noise unless 
addressed at the galaxy level and 
statistically \cite{dawson2016, Mandelbaum2018}.  If blends can be tagged 
(for example via a color gradient in the combined core) then they may be 
eliminated from the sample. Conspicuous blends with larger separation may be
more easily identified, tagged and deblended.

Another challenge for WL is intrinsic alignment 
\cite{catelan2001,hirata2004}. The orientation of a galaxy can be 
influenced by the larger structure containing it, thus breaking the 
assumption that galaxies orient randomly in the absence of lensing.
The correlation of galaxy orientations in the same structure would
contaminate the true shear signal arising from these galaxies' common 
foreground mass distribution. Moreover, if these galaxies themselves 
are taken as the foreground, then the structure they reside in would 
lens galaxies in the background. The two effects of the same structure, 
intrinsic alignment of the foreground galaxies and shear on the 
background galaxies, would correlate the 
background and foreground galaxy orientations, contaminating
the true correlation between these galaxies' shear signals. 
One needs to utilize \phz{} information and model the intrinsic 
alignment effects to minimize the impact on cosmology
\cite{king2002,king2005,joachimi2010,heymans2013}. It is encouraging
that marginalization of the intrinsic alignment systematics using
a simple phenomenological model may be sufficient for shear 
analyses \cite{troxel2015,krause2017}. 
For the LSST, faint blue galaxies will dominate its gold sample, 
and these are expected to be less affected by intrinsic alignment 
\cite{mandelbaum2011}.

Since magnification is extracted from the observed galaxy density 
fluctuations, it avoids some of the difficulties with shear 
measurements but unfortunately inherits its own systematics. 
The galaxy overdensity with magnification effect
is approximately \cite{broadhurst1995,moessner1998}
\beq
\delta^{\tilde{g}} \simeq b \delta+5(s-0.4)\kappa,
\eeq
where $s = \rmd \log N/\rmd m\sim 0.2$--$0.6$ is the logarithmic slope 
of the galaxy number counts at the limiting magnitude $m$ of the 
sample \cite[e.g.,][]{tyson1988}.
The total angular power spectrum of galaxies in \phz{} bins $i$ 
and $j$ becomes \cite{yang2015}
\beq \label{eq:gaps-tot}
C_{ij}^{\tilde{g}\tilde{g}} = C_{ij}^{gg} + q_i C_{ij}^{\kappa g} + 
q_j C_{ij}^{g\kappa} + q_i q_j C_{ij}^{\kappa\kappa},
\eeq
where $C_{ij}^{gg}$ is the power spectrum due to galaxy intrinsic 
clustering, i.e., \eref{eq:gaps} with the superscript $g$ written
out explicitly, $q_i = 5(s_i-0.4)$ with $s_i$ for bin $i$,
$q_i C_{ij}^{\kappa g}$ and $q_j C_{ij}^{g\kappa}$ are the galaxy-magnification 
cross power spectra, and $q_i q_j C_{ij}^{\kappa\kappa}$ is the magnification 
power spectrum. Since the convergence for a lower redshift bin is 
uncorrelated with the galaxy overdensity in a higher redshift bin, 
one of the cross terms vanishes for two well separated bins.
The galaxy-convergence power spectra $C_{ij}^{\kappa g}$ 
and $C_{ij}^{g\kappa}$ and the convergence power spectrum 
$C_{ij}^{\kappa\kappa}$ can be calculated using 
\eref{eq:saps} but with the two shear window functions replaced 
by the galaxy window function \eref{eq:Wg} and the convergence window 
function, which is identical to the shear window function \eref{eq:Ws}.

The intrinsic clustering term is several orders of magnitude greater 
than the rest in \eref{eq:gaps-tot} when the \phz{} bins $i$ and $j$ 
overlap significantly in true-redshift space; even if they are 
well separated, the cross term $C_{ij}^{g\kappa}$
--- assuming that bin $j$ is at higher redshift --- still overwhelms
$C_{ij}^{\kappa\kappa}$ \cite{yang2011}. Furthermore, galaxies can be 
assigned to wrong redshift bins because of \phz{} errors, and the resulting 
spurious correlations can be a serious contaminant for the 
magnification signal (conversely, magnification in tomographic galaxy
cross-correlations may be a contaminant for \phz{} self-calibration 
with the cross-correlations). Magnification systematic error can also 
arise from exposure depth spatial variations, dust, and blending of galaxy 
images \cite{Morrison2012}. Therefore, like the case with shear, care must 
be taken when utilizing lensing magnification for precision cosmology. 
As a cosmological constraint, magnification is subdominant to the other WL 
probes.

\subsection{Type Ia supernovae}
SN Ia distances provided the most direct evidence for the accelerated 
cosmic expansion \cite{riess98,perlmutter99} and will remain an important 
dark energy probe in the future, though systematic errors make it a
less precise probe of cosmology by comparison to a joint analysis of WL 
and BAO.  The LSST is expected to obtain 
more than ten thousand well observed SNe Ia ($0.1<z<1.2$) in the 
deep-drilling fields, hundreds of thousands reasonably-well observed 
SNe Ia ($z \lesssim 0.8$) in the main survey, and millions of 
detection-only SNe Ia \cite{lsst_sb}. 
Such a huge sample of SNe Ia will enable new tests of cosmology such as 
the isotropy of distance-redshift relation over the whole survey area. 

The utility of SNe Ia for cosmology is based on their standardizable 
peak luminosity that is tightly correlated with the initial decline 
rate of the light curve \cite{phillips1993}. 
As such, one can obtain the distance modulus, which
is a function of cosmological parameters, via 
\cite{jha2007,guy2007,conley2008}
\beq
\mu_B = m_{B} - M_B + \alpha x_1 - \beta \mathcal{C}
\label{eq_sn_mu}
\eeq
where $m_B$ is the rest-frame \emph{B}-band peak apparent magnitude, 
$M_B$ is the 
corresponding absolute magnitude, $x_1$ is the shape parameter of the
light curve, $\mathcal{C}$ is the $B-V$ color, and $\alpha$ and $\beta$
are nuisance parameters to be fitted along with $M_B$ (actually 
$M_B-5\log h$) and the cosmological parameters. Given the fairly low 
intrinsic dispersion of $\mu_B$ of roughly $0.12$, a single SN Ia can 
provide a luminosity distance with a nominal error of $6\%$ (excluding 
the error in $h$). It is worth mentioning 
that SNe Ia are likely to be more standard in the rest-frame near infrared 
\cite{wood-vasey2008,barone-nugent2012}, though the LSST can observe only up
to $1.05\um$.

The degeneracy between the Hubble constant and $M_B$ removes the absolute
scale of SN Ia distances and thus weakens their constraints on the cosmological
parameters. The remedy is to obtain a sample spanning as wide a redshift 
range as possible and to combine a tight prior on the Hubble constant from 
external datasets. A number of technical 
issues and systematic uncertainties could be the limiting factors for 
SN Ia cosmology with next-generation surveys. Besides the challenges of
obtaining very well sampled light curves and precisely calibrated 
photometries, one also needs further investigation on potential 
evolution of SN Ia properties over redshift, the relation 
between the SN Ia peak luminosity and its host properties, 
and certainly the physics of SN Ia explosions.

The LSST survey, especially its deep-drilling fields, is tasked to 
address the technical challenges for SN Ia cosmology, though effort is 
still being made to optimize the operation strategy of the main survey. 
The huge LSST SN Ia sample can be divided into many sub-samples for 
detailed studies of various systematics intrinsic to SNe Ia themselves as 
well as biases arising from sample selection. Nonetheless, 
being an imaging survey, the LSST has its own challenges. For example, it 
has to correctly identify SNe Ia and determine their redshifts based on
photometric data only \cite{sullivan2006,sako2011}; 
even though a small fraction of them will  
have spectroscopic follow-ups by other facilities, the LSST still has to 
send out alerts promptly for the best candidates among millions of 
transients every night with very high success-rate. It is also noted
that \phz{} errors increase the apparent dispersion of the 
standardized SN Ia peak luminosity and that the sample-averaged 
distance modulus--\phz{} relation departs from the distance modulus--true 
redshift relation \cite{zentner2009}. One should account for these 
effects when estimating the cosmological parameters. 

\subsection{Galaxy clusters}
Galaxy clusters are the largest gravitationally bound systems in the universe.
At the high-mass end, a cluster may have thousands of member galaxies, 
while a low-mass cluster can still contain many dozens of galaxies.
Cluster cosmology is a rich topic and involves observations from microwave
to X-ray \cite[for a review, see][]{allen2011}. The primary statistic of 
cluster cosmology for the LSST is the mass function, i.e., number 
density of clusters as a function of mass at a certain redshift. 
Since clusters have grown under 
gravitational instability from relatively high density regions of 
the fairly Gaussian initial density fluctuations, one can roughly estimate 
the cluster mass function based on the statistics of 
the Gaussian random field \cite{press1974,bond1991}. 
For precision cosmology, more accurate mass functions need to be 
extracted from \emph{N}-body simulations \cite{jenkins2001,tinker2008}. 

The cluster mass function is sensitive to both the geometry of the universe 
and the growth of the large-scale structures, which makes it a 
powerful tool \cite{haiman2001,majumdar2004,rozo2010}. 
However, the utility of clusters for precision cosmology critically depends 
on the knowledge of the mass-observable relation and its scatter 
\cite{lima2005}. Because of the highly nonlinear process 
of cluster formation and evolution, one has to rely on state-of-the-art 
simulations and observations to precisely calibrate the mass-observable 
relation and its scatter for a wide range of cosmology. 
It takes considerably more effort to 
properly include all the relevant physics such as star formation and feedback 
in the simulations to reproduce the observables
\cite[e.g.,][]{frenk1999,nagai2007,kravtsov2012}.

The LSST will produce a well observed sample 
of several hundred thousand clusters, much larger than any previous surveys.
It is not trivial to construct such a sample efficiently from the LSST 
photometry data with high purity, high completeness, and well understood 
selection function. Accurate \phz{}s are needed for clusters as well. 
Averaging over member galaxies' redshifts generally reduces the \phz{} 
errors, but one has to be careful about any systematic redshift errors 
of the galaxy population in the same cluster. 
Although WL measurements from the same LSST data can provide 
important absolute cluster mass calibration, the scatter will be rather 
high ($\gtrsim 20\%$), mainly due to triaxiality of the halos, projection 
effects along the line-of-sight, and shape noise of shear measurements 
in the limited FoV
\cite{becker2011}. It has been shown for Dark Energy Survey (DES) that 
even a small fraction of the whole cluster sample (e.g., 200 out of $10^5$ 
expected) calibrated against a mass proxy of $10\%$ scatter can enhance 
the DES dark energy constraints by $50\%$ \cite{wu2010}. 
As such, it is crucial to establish a tight relation 
between cluster observables from LSST data with the cluster mass through 
external low-scatter mass proxies, likely from X-ray 
observations \cite{mantz2010b}. 

\subsection{Strong lensing}

While WL effects are hardly discernible to the eye, background
objects can be stretched into arcs or even Einstein rings and be split into 
multiple images in the strong lensing regime 
\cite[e.g.,][]{colley1996,inada2003,king1998}. 
A lens can be a galaxy or a galaxy cluster, and the background sources are 
usually galaxies that are effectively static within the time span of an 
observational program. In rare cases, one can find multiple-image 
systems of variable or transient background sources such as active 
galactic nuclei (AGNs) and SNe \cite{kelly2015,goobar2017}. 
Different images of the same source means 
different paths and travel times for the photons. The detailed configuration
of the images and time delay between them are determined by the mass
distribution of the lens and the angular diameter distances between the 
observer, the lens, and the source. Therefore, the strong-lensing time delay 
can be a useful probe of cosmology. It is most sensitive to the Hubble 
constant \cite{refsdal1964} and has recently been demonstrated to be able
to constrain other cosmological parameters such as the curvature parameter 
$\Ok$ and the dark energy EoS parameter $w$ \cite{suyu2010,linder2011}.

The typical time delay of a double lens (double-image system) is 1-3 
months, and much shorter delays of less than a day can occur when two 
images in a quad lens (quadruple-image system) are very close to each 
other \cite{oguri2010}. 
The LSST's 10-year rolling survey with 3-month seasons and a cadence of 5 
days is suited for time delay measurements of typical strong lensing 
systems. It is estimated that the LSST strong lens sample will be at 
least an order of magnitude larger than previously obtained, allowing 
time delays to be measured for more than 3000 strongly lensed quasars 
and more than 100 strongly lensed SNe \cite{oguri2010}. 

Like other LSST probes, strong-lensing time delay has its own challenges.
Besides lens monitoring and fitting for the time delay, one would also 
need accurate information about the lens mass distribution and 
the source redshift. Moreover, the impact of line-of-sight
mass distribution on time delay needs to be taken into account, 
at least statistically. In these areas, high-resolution observations from
space, deep spectroscopy on the ground, and modeling of the foreground
structures will be particularly helpful for strong-lensing cosmology with 
the LSST. In fact, with a major undertaking of external follow-up observations,
one can also probe the geometry of the universe with strongly lensed sources 
in multiple source planes \cite{gavazzi2008,jullo2010,collett2014}.

\subsection{Joint analyses}

The ability to produce large uniform data sets with high quality 
for multiple cosmological probes is a crucial advantage of the LSST. 
These probes are affected by various systematics and are sensitive to 
cosmology in different ways. Joint analyses of multiple probes not only 
improve parameter constraints but, more importantly, also enable 
certain cross-calibrations of known systematics and detection of 
unknown systematics. This allows one to perform robust tests on  
cosmological models and to explore with confidence fundamental physics of the universe 
beyond current understanding.

The aforementioned probes are not always statistically independent. 
Correlations arise when two probes are linked by common factors of real
physical origin or observational effects. 
For instance, WL shear of background galaxies is essentially 
a function of the weighted projection of the line-of-sight mass 
distribution, which is traced by galaxies. Thus, foreground galaxy 
density fluctuations are correlated with the background shear, 
similar to the case of lensing magnification in \sref{sec:wl}. 
This effect is known as the galaxy-shear correlation or
galaxy-galaxy lensing on large scales ($\ell \lesssim 1000$)
\cite{hu04a,zhan06b,bernstein2009,mandelbaum2013}, though, strictly 
speaking, the former can also include a contribution from intrinsic 
alignment. The fact that WL and LSS (or BAO) techniques share, 
at least partially, the same 
catalog of galaxies, and their \phz{} systematics also induces
a correlation between the two probes. Such a correlation can be
beneficial, as the self-calibration of the \phz{} error distribution by 
cross-bin galaxy power spectra can reduce the sensitivity of
the joint cosmological constraints of WL and BAO to 
uncertainties in the photo-z error distribution 
\cite{zhan06b,schneider2006,zhang2010}.  Conversely, because of
the shared dark matter structure, the complementary WL and BAO 
measurements break degeneracies in either probe.

\begin{figure}
\centering
\includegraphics[width=0.9\columnwidth]{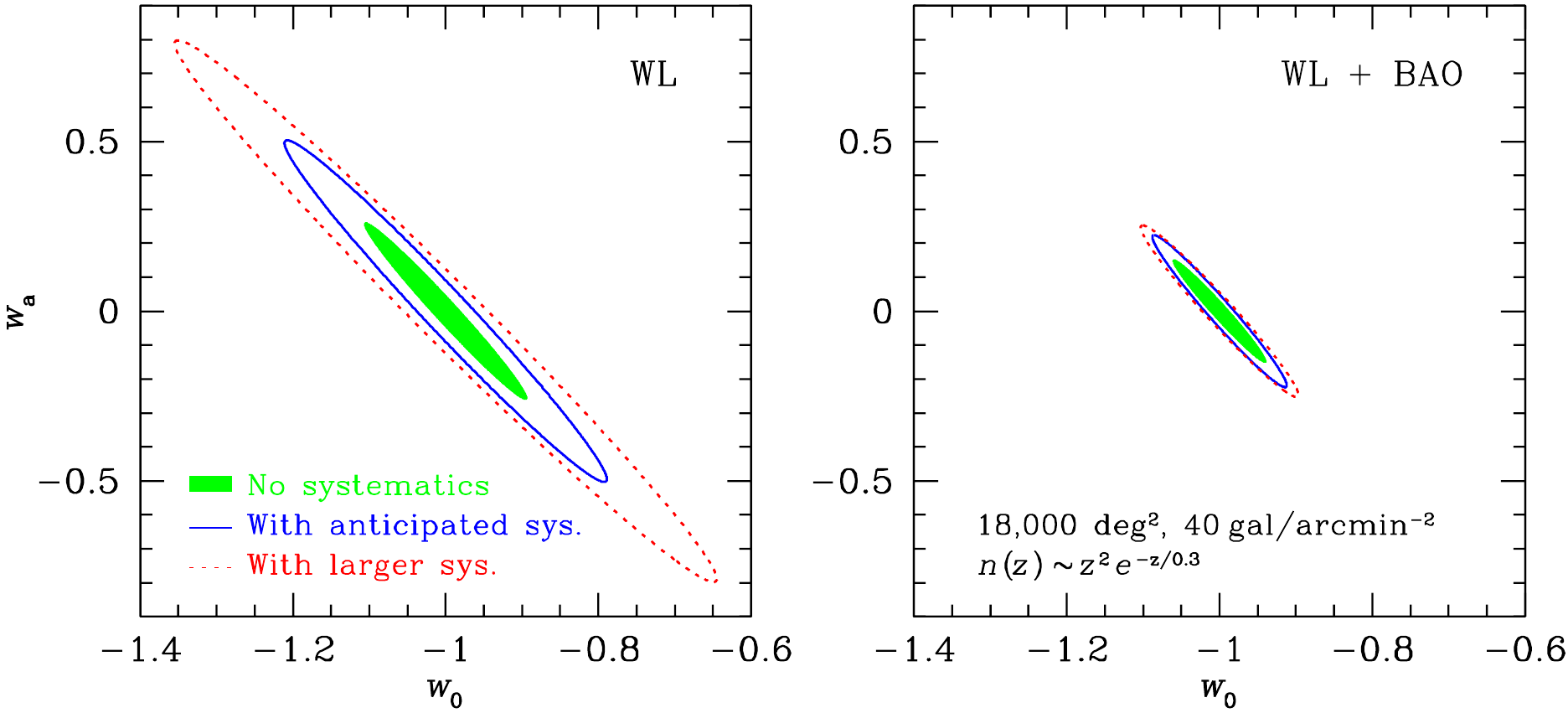}
\caption{One-sigma error contours of the dark energy EoS parameters
$w_0$ and $w_a$ from LSST WL shear power spectra alone 
(left panel) and shear and galaxy power spectra analyzed jointly 
(right panel). The shaded areas represent the results with 
statistical errors only. The solid contours correspond to those
with the anticipated level of systematic errors, which include 
the uncertainty in the \phz{} error distribution and additive and 
multiplicative errors in the power spectra. 
The assumed \phz{} systematics would require a redshift calibration 
sample of 3000 spectra per unit redshift interval if the \phz{} 
error distribution were Gaussian. This calibration can be achieved 
statistically by using spectra of the brighter galaxies in the 
redshift interval in small areas of the sky. The dotted contours 
relax the requirement to 188 spectra per unit redshift. Figure adapted 
from \cite{zhan2009b} with updated survey data model.
\label{fig:wsys}}
\end{figure}

\Fref{fig:wsys} demonstrates the synergy between WL and BAO
based on Fisher matrix analyses \cite{jungman1996,vogeley1996,tegmark1997}. 
It is adapted from \cite{zhan2009b} with the survey area, galaxy 
redshift distribution, and galaxy surface number density, respectively,
adjusted to 18,000$\deg^2$, $n(z) \sim z^2\exp(-z/0.3)$, and 
$40\arcmin^{-2}$. The number density is an effective value for WL
measurements, and the same is applied to BAO for convenience. 
The constraints of the joint analysis are derived from the full set 
of galaxy-galaxy (i.e., galaxy clustering), galaxy-shear 
(i.e., galaxy-galaxy lensing), and shear-shear power spectra. 
These power spectra are affected by different powers of the galaxy 
bias and together can achieve robust cosmological constraints in the
presence of uncertainties in the galaxy bias \cite{hu04a,zhan06b}. 
Numerically, the galaxy-shear power spectra are 
equal to the galaxy-convergence power spectra $C_{ij}^{g\kappa}$ in 
\sref{sec:wl}. \emph{Planck} CMB priors from a Fisher matrix 
calculation in \cite{zhan06b} are applied to all the results.
While the constraints on the dark energy EoS parameters, 
$w_0$ and $w_a$, from shear power spectra alone are sensitive to 
systematic uncertainties in the \phz{} error distribution, the joint 
results of shear and galaxy power spectra remain fairly immune to these 
systematics. The dramatic improvement of the joint results over the 
WL-only results is due to the mutual calibration 
of the \phz{} uncertainties as well as the uncertain factor of the 
galaxy bias. In other words, much of the complementarity is in 
parameter space that has been marginalized over. 
A demonstration of the constraining power of this joint analysis of cosmic shear, 
galaxy-galaxy lensing and angular clustering has recently been obtained
\cite{vanUitert2017}.

\begin{figure}
\centering
\includegraphics[width=\columnwidth]{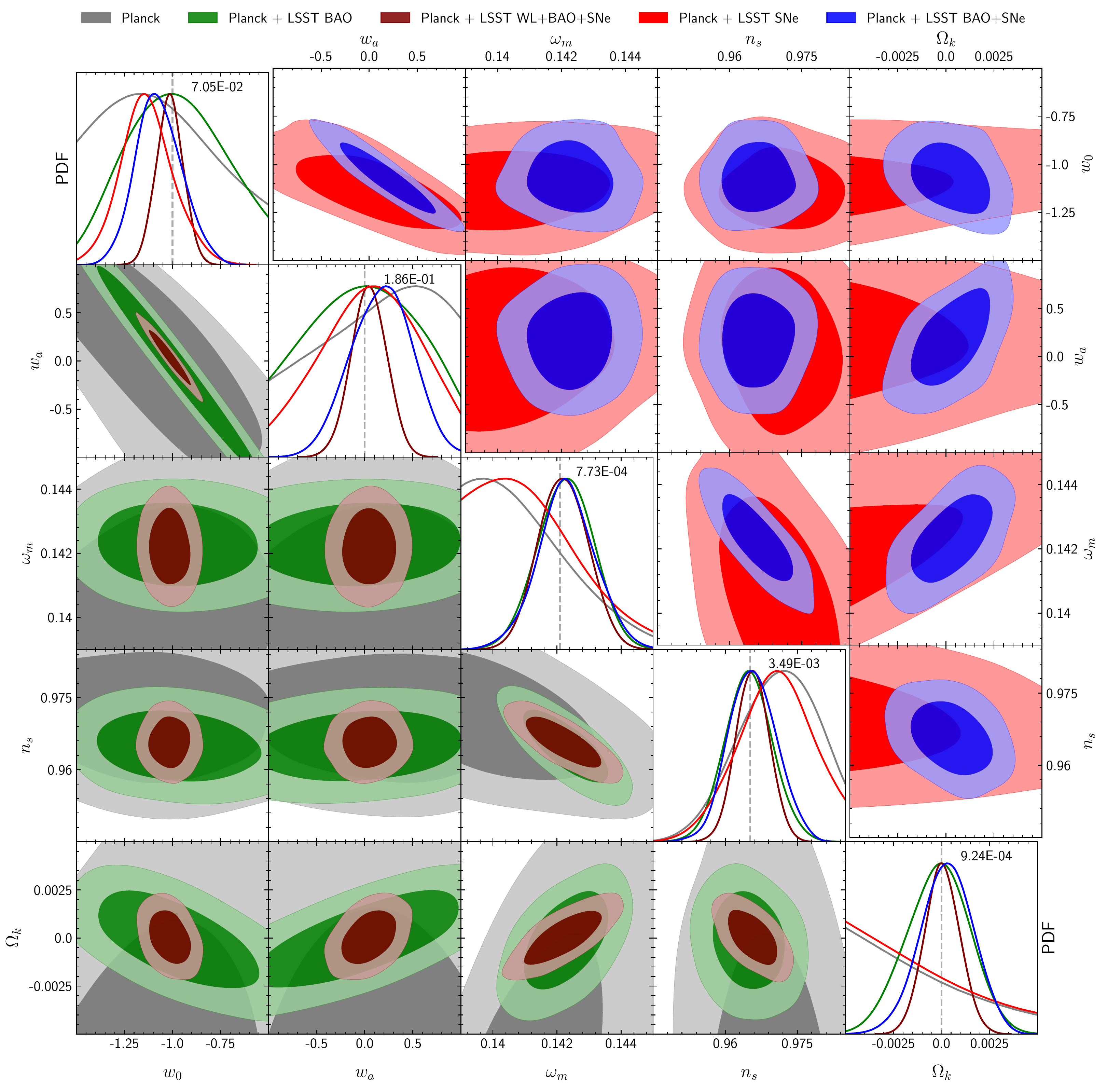}
\caption{Forecast of
cosmological constraints from combinations of three LSST probes 
and real \emph{Planck} data. For a proper layout, only 5 parameters
that benefit most from the LSST data are plotted. 
The probes are color-coded: \emph{Planck}-only in gray, 
LSST SNe Ia plus \emph{Planck} in red, LSST BAO plus \emph{Planck} in green, 
LSST BAO+SNe Ia plus \emph{Planck} in blue, and LSST WL+BAO+SNe Ia
plus \emph{Planck} in brown. 
The shades of the same color in each two-parameter panel fill the
contours of $68\%$ and $95\%$ confidence level, respectively. 
The panels along the diagonal present the marginalized probability 
distribution functions of each parameter, which are normalized to have a 
maximum value of unity. The vertical dashed lines in the panels along the 
diagonal mark the parameters' fiducial values, and the 1-$\sigma$ errors 
of LSST WL+BAO+SNe Ia plus \emph{Planck} are also given in these panels.
\label{fig:cswb}}
\end{figure}

A forecast of cosmological constraints from combinations of three 
LSST probes and real \emph{Planck} data is given in \fref{fig:cswb}. 
The \emph{Planck} data consists of the joint likelihood of the 
temperature-temperature (TT), temperature-E polarization (TE), 
E polarization-E polarization (EE), and B polarization-B polarization
power spectra in the range of $2 \le \ell \le 29$ and the 
foreground-marginalized joint likelihood of TT ($30 \le \ell \le 2508$), 
TE ($30 \le \ell \le 1996$), and EE ($30 \le \ell \le 1996$) band power 
spectra \cite{planck2016}. The CMB power spectra are calculated using
{\sc class} with the lensing effect turned on \cite{blas11}. 
The LSST probes considered are WL, BAO, and SNe Ia. 
The set of cosmological parameters includes
$w_0$, $w_a$, $\om$, $\ob$, $\Ok$, $n_\mathrm{s}$, $H_0$, the redshift of
reionization $z_\mathrm{rei}$, the helium fraction $Y_\mathrm{P}$, and 
the amplitude of the curvature power spectrum $A_\mathrm{s}$. The sum of 
neutrino masses is fixed at $0.06\ev$. 

The posterior probability distribution of the parameters are mapped 
with an affine invariant Markov Chain Monte Carlo sampler 
\cite{goodman2010,xu2017}. For LSST WL and BAO, we use the corresponding Fisher 
matrices to calculate contributions to the final likelihood. The Fisher
matrices have been marginalized over non-cosmological parameters such as 
those modeling the \phz{} errors, galaxy biases, and errors in the power 
spectra. The fiducial model of the Fisher matrices coincides with the 
\emph{Planck}-only best-fit \lcdm{} model \cite{planck2015par}, 
and the fiducial values of 
the additional parameters, $w_0$, $w_a$, and $\Ok$, are set to 
$-1$, $0$, and $0$, respectively. For LSST SNe Ia, a simple mock 
sample (tuples of distance modulus and redshift) is generated under 
the fiducial cosmological model according to the yields and redshift 
distributions from section 11.2 of \cite{lsst_sb}. 
The selection criterion is that each SN must be observed in three or 
more filters at a minimum signal-to-noise ratio of 20, which results in 
roughly 400,000 SNe from the main survey and 12,000 SNe from the 
deep-drilling survey in 10 years. 
While the deep-drilling survey can be optimized for SN observations, it 
is impossible for the main survey to mass-produce SN light curves with 
the same quality as can be achieved by the former. To be conservative, 
we reduce the number of SNe Ia in the main sample somewhat arbitrarily
to 100,000. The intrinsic dispersion of the standardized SN Ia peak 
absolute magnitude is assumed to be $0.13$, and we adopt a \phz{} rms 
error model of $\sigma_z=0.02(1+z)$ \cite{palanque2010,kessler2010,wang2015}.
\Phz{} bias errors and correlations between different SNe are neglected 
for simplicity. The likelihood of the SNe Ia is then evaluated in the 
conventional way \citep[see, e.g.,][]{astier2006} with modifications to 
the fiducial distance modulus-redshift relation and apparent dispersion 
of the distance modulus to account for the effects of 
the \phz{} errors \cite{zentner2009}. 

\Fref{fig:cswb} shows that the LSST is indeed more than a dark energy 
experiment. With the huge number of modes contained in the 
survey volume \cite{zhan06a}, LSST BAO is able to achieve significant
enhancement over \emph{Planck} on parameters that affect the shape of 
the matter power spectrum. The CMB can only determine
one distance --- the distance to the last scattering surface, so that
\emph{Planck} alone, even with the CMB lensing effect, does not provide 
useful constraints on $w_0$, $w_a$, and $\Ok$. 
Moreover, the curvature parameter is pulled slightly negative:
$\Ok = -0.018_{-0.009}^{+0.014}$. Similar behavior is observed by the 
\emph{Planck} Collaboration as well when the \lcdm{} model is extended 
to include $\Ok$ \cite{planck2015par}.
The addition of LSST SNe Ia to \emph{Planck} is particularly helpful 
on $w_0$ and, to a lesser degree, $w_a$, but, as expected, not much is 
gained on the other parameters. The joint analysis of 
LSST WL+BAO+SNe Ia plus 
\emph{Planck} places tight constraints on the dark energy 
EoS parameters, obtaining $\sigma(w_0)=0.07$, $\sigma(w_a)=0.19$, and 
the minimum error of the EoS $\sigma(w)=0.017$ at the pivot redshift 
$z=0.58$, and reduces the uncertainties on the other 
parameters moderately over those of LSST BAO plus \emph{Planck}. 
Although not shown in \fref{fig:cswb}, the joint results of LSST WL+BAO 
plus \emph{Planck} are nearly identical to the all-probes 
results except that the errors on $w_0$ and $w_a$ are roughly $10\%$ larger.

Since we have not included all the LSST cosmological probes in the 
forecast, there is still room for improvement on the constraints. 
Efforts are being made by the LSST DESC to develop a suite of 
software packages for comprehensive analyses of the data
\cite{desc_wp,krause2017}. 
Moreover, the $w_0$-$w_a$ parametrization does not capture 
the full complexity of all dark energy models and significantly 
underestimates the capabilities of Stage IV surveys like the LSST 
\cite{albrecht07,albrecht2009}. 
It has been demonstrated that multiple dark energy 
EoS eigenmodes in redshift (or expansion factor) space can be 
reconstructed from the data with sufficiently small errors to allow tests 
of a wide variety of dark energy models \cite[e.g.,][]{barnard08}.
Finally, anisotropy in dark energy introduced, for example, by dark energy 
clustering can be detected in the LSST survey via two complementary methods 
\cite[section 15.5]{lsst_sb}.

\section{Summary} \label{sec:summary}

The LSST holds great promise for breakthrough discoveries in cosmology, 
and, not surprisingly, it comes with challenges. 
While the LSST project works on the technological and engineering 
front, scientists have much to prepare for. On the one hand, the LSST 
survey will bring unprecedented statistical precision enabling 
knowledge beyond our current horizon, requiring extraordinary effort 
and often novel use of the data.
On the other hand, like other large-scale surveys in the 2020s, 
the LSST is likely to remain limited by systematic errors arising
from instruments, observations, data reduction, analysis methods, and
even deficiencies in theories. We have largely skated over these 
important issues to avoid distraction by the details. Efforts on 
detecting and controlling systematics, especially those not yet known, 
are crucial for the LSST to reach for its full potential. In fact,
a large fraction of the activities of the LSST DESC is devoted to
systematics.

Much of the science produced by the LSST survey would greatly benefit 
from other data at a variety of 
wavelengths, resolutions, depths, and timescales. 
The \emph{Euclid} mission \cite{euclid} and 
the Wide Field Infrared Survey Telescope (\emph{WFIRST}) \cite{wfirst} 
can provide near-infrared photometry data in the area partially
overlapping with the LSST survey.
The galaxy spectral energy distributions derived from 
the combined data should give rise to even better photometric 
redshifts as well as tighter constraints on stellar masses and star 
formation histories crucial for galaxy evolution studies.  
The WL analyses from space and from the ground will 
be highly complementary and will provide cross-checks of one another. 
More importantly, in a subset of the survey area, \emph{Euclid}
and \emph{WFIRST} spectra of galaxies 
in the bright part of the LSST photometric sample will efficiently 
calibrate the LSST \phz{} error distributions in redshift bins. 

The LSST will also enable multi-wavelength studies of faint optical
sources using gamma-ray, X-ray, infrared and radio data. The LSST will provide 
a crucial complementary capability to space experiments operating in 
other wavebands, such as the ongoing Nuclear Spectroscopic Telescope 
Array \cite{nustar} and 
the {\it Fermi} Gamma-ray Space Telescope \cite[e.g.,][]{fermi_lat}. 
The large samples of various astronomical source populations 
will yield both representative objects of each population with 
exquisite statistics and extremely rare objects for investigations by 
powerful facilities such as the James Webb Space Telescope \cite{jwst} 
and the next generation of $30$--$40\m$ telescopes \cite{gmt,tmt,e-elt}.  
Better understanding of these sources, especially on the subject of 
galaxy evolution, will in turn aid the cosmological data analysis.
The Square Kilometer Array \cite{ska} will extend the cross calibration
using a large sample of galaxies in the radio.

There is no doubt that close collaborations between the LSST and its 
peers will make ``the whole greater than the sum of the parts'' 
\cite{lsst+euclid+wfirst}. However, with other surveys, the full 
benefit is likely to require joint reduction and analyses of the 
data at fairly low levels, e.g., images as opposed to catalogs; 
in other cases, concerted observational campaigns of serious 
undertaking may be necessary. 

The LSST will be transformative 
in many areas beyond astronomy and physics as sciences and beyond 
astronomy and physics as communities. 
The most exciting aspect of the LSST is the enormous 
discovery space its offers.  What we have imagined of the LSST today 
may well be only the tip of the iceberg when the survey unfolds. 

\ack

Financial support for the LSST comes from the NSF through Cooperative 
Agreement No. 1258333, the Department of Energy Office of Science 
under Contract No. DE-AC02-76SF00515, and private funding raised 
by the LSST Corporation. HZ was partially supported by the National
key Research Program of China grant No. 2016YFB1000605.
We thank Youhua Xu for assistance in producing the figures,
and Michael Schneider, Seth Digel, Zeljko Ivezic and the anonymous
referee for helpful suggestions for the manuscript.

\bibliographystyle{iopart-num}

\small

\providecommand{\newblock}{}

\end{document}